\documentclass[twocolumn]{aastex631}
\usepackage{amsmath}
\usepackage{booktabs}

\newcommand{\gname}{G\,11.2–0.3}

\newcommand{\uv}{\textit{u-v}}

\newcommand{\JyPerBeam}{$\,{\rm Jy\,beam^{-1}}$}
\newcommand{\mJyPerBeam}{$\,{\rm mJy\,beam^{-1}}$}

\newcommand{\cm}{\,cm}

\begin{document}

\title{Radio Observation of the Pulsar Wind Nebula in SNR G11.2$-$0.3}

\author{Yu Zhang}
\affiliation{School of Physics and Astronomy, Sun Yat-Sen University, No. 2 Daxue Road, 519082, Zhuhai China}

\author{Yihan Liu}
\affiliation{School of Physics and Astronomy, Sun Yat-Sen University, No. 2 Daxue Road, 519082, Zhuhai China}
\affiliation{Department of Physics, The University of Hong Kong, Pokfulam Road, Hong Kong}

\author[0000-0002-5847-2612]{C.-Y. Ng}
\affiliation{Department of Physics, The University of Hong Kong, Pokfulam Road, Hong Kong}

\author{Mallory S. E. Roberts}
\affiliation{Eureka Scientific, 2452 Delmer Street Suite 100, Oakland, CA 94602, USA}

\author{Lili Yang}
\affiliation{School of Physics and Astronomy, Sun Yat-Sen University, No. 2 Daxue Road, 519082, Zhuhai China}
\affiliation{Centre for Astro-Particle Physics, University of Johannesburg, P.O. Box 524, Auckland Park 2006, South Africa}

\correspondingauthor{Yihan Liu}
\email{liuyh363@mail.sysu.edu.cn}

\begin{abstract}

Pulsar wind nebulae (PWNe) are important sources for understanding galactic high-energy processes, but it is controversial until now about how high-energy particles in PWNe are accelerated and transported.
Lacking radio counterparts of X-ray PWNe (the proposed acceleration sites) introduce difficulties to better understandings in multi wavelengths.
Our recent 3, 6, and 16\,cm high-resolution observations of G11.2$-$0.3 PWN with the Australia Telescope Compact Array (ATCA) uniquely show morphological similarity with its X-ray PWN (a torus/jet feature). Spectral indices of the radio torus and jet are around -0.09 and -0.10, respectively. Meanwhile for the jet region, the spectral break between radio and X-ray spectra implies particle acceleration mechanisms other than a diffusive shock acceleration. Polarization results suggest a helical B-field inside the jet, the equipartition B-field strength of which is below 100\,$\mu$G.

\end{abstract}


\section{Introduction}
\label{sec:intro}

After a massive progenitor star undergoes a supernova (SN) explosion, it may leave behind a compact neutron star inside of a supernova remnant (SNR), which may also be observed as a pulsar, emitting periodic signals towards the Earth due to its rotation. The slowing of the pulsar's rotation powers this pulsed emission as well as an outflow of particles in a pulsar wind (PW). This wind inflates a magnetic nebula full of accelerated high-energy particles inside, forming a pulsar wind nebula (PWN). 
The two main emission mechanisms from injected energetic particles in PWNe are synchrotron (SYN) emission and inverse Compton (IC) emission. Relativistic electrons spiral along magnetic field lines to generate non-thermal SYN emission from radio to hard X-rays.  These same particles can also interact with low-energy photons to radiate IC emission in gamma-ray bands \citep[see][]{2017hsn..book.2159S}. 
Recent observations indicate that PWNe should be a significant fraction of galactic leptonic very-high-energy (VHE) sources. For instance, many PWNe sources are found associated with LHAASO or H.E.S.S. sources, with observed photon energies up to 1\,PeV coming from the Crab Nebula \citep{2021Natur.594...33C} and other PWNe.
How these high energy particles are accelerated and transported, as well as the electron and positron components of cosmic rays are crucial questions to understand. However, observations have still not successfully discriminated between different acceleration models \citep{2011ApJ...726...75S,giacinti2019electronaccelerationcrabnebula}.

\begin{figure}[h!]
    \centering
    \includegraphics[width=0.8\linewidth]{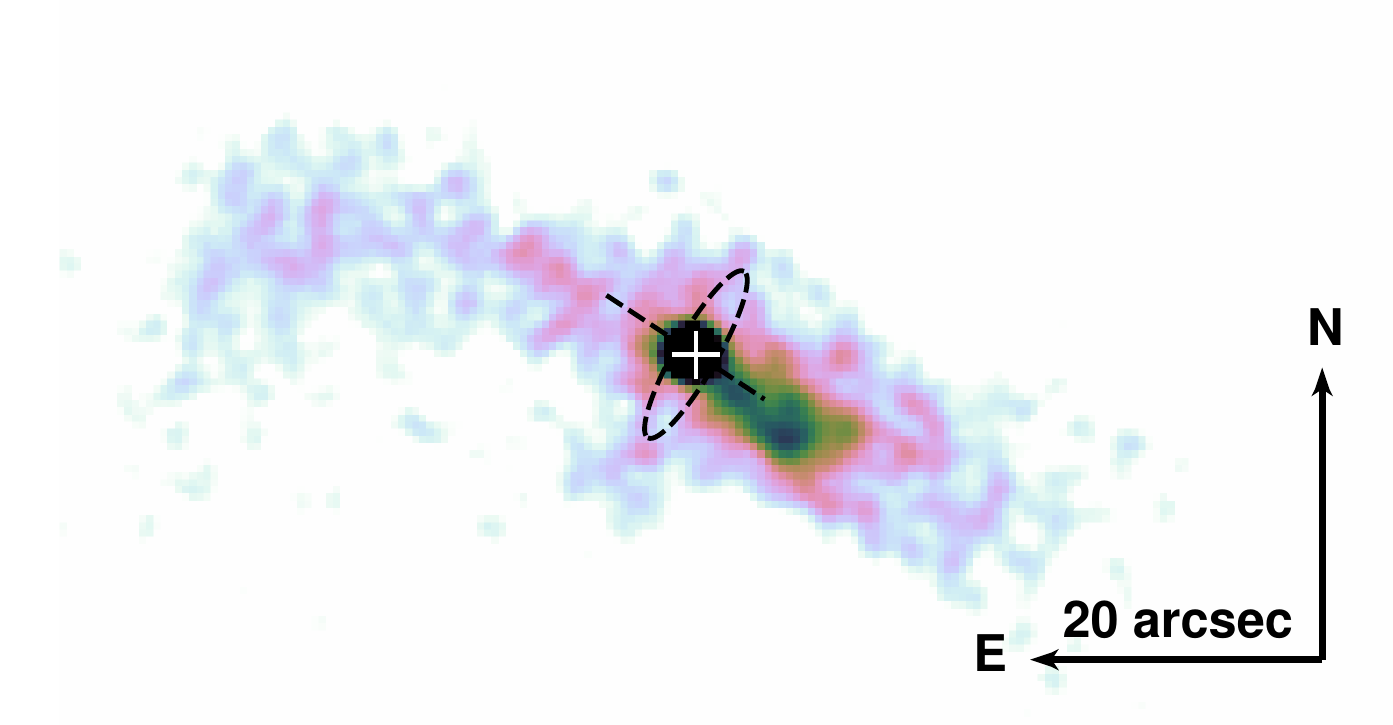}
    \caption{ 4-9\,keV \emph{Chandra} image of the \gname\ PWN. The dashed line and ellipse suggest torus and jet structure of this PWN; the cross indicates to the pulsar position.}
    \label{fig:jet_torus_g11}
\end{figure}

Even though diffusive shock acceleration (DSA) is popular to explain how particles gain high energies in PWNe, it is difficult to understand why the Crab PWN shows multiple breaks in its power-law spectrum from radio energies to soft $\gamma - $rays. This is further complicated by  low energy emission showing different morphologies from higher energy X-ray structures \citep{1999A&A...346L..49A,2001ApJ...560..254B,2017ApJ...840...82D}. 
One leading alternative model suggests that particles in the striped PW could experience magnetic reconnection and then the reconnected magnetic field will transfer magnetic energy to particle energies, this can also be a way to explain why magnetization is required to be high near the pulsar but low just beyond the wind termination shock (TS) region \citep{2011ApJ...726...75S}. 
A typical reconnection model requires a broken power law distribution of the particle energy (as well as SYN spectrum), rather than a simple power law that could be formed in the scenario of fermi acceleration. 
Related radio observations are therefore crucial to clearly pin down the synchrotron spectrum and reveal the frequency regime of the spectral break between the radio and X-ray bands. 

We note that significant synchrotron cooling could also induce a break in the SYN spectrum; to exclude this possibility, it is wise to consider particles accelerated just beyond the PWN termination shock (i.e., in the X-ray structures).  However, if the radio emitting electrons are cooled to produce the X-ray emitting particles near the TS, then that just transfers the problem to the X-ray regime, since there is clear evidence of X-ray softening as the wind moves away from the TS in many PWNe, which is usually attributed to synchrotron cooling \citep{2023ApJ...945...82L,2008ApJ...684..542K,2016ApJ...820..100M,2006ARA&A..44...17G}. 

High-resolution X-ray observations have revealed a common geometry of an equatorial torus with jets along the spin axis for many young PWNe \citep{2004ApJ...601..479N,2008ApJ...673..411N,2008AIPC..983..171K}, which can be explained with a latitude-dependent intensity of pulsar winds \citep{2002AstL...28..373B}. 
In addition, other X-ray structure have potential connections with different particle acceleration mechanisms and sites in the nebula. For instance, the inner knots and dynamical wisps of the Crab Nebula suggest the.transportation of relativistic particle streams insidethe nebula. There are also interesting loops possibly formed via the kink instability in high energy structures \citep{2004ApJ...616..403S,2017ApJ...840...82D}. 
However, one difficulty is that X-ray PWN counterparts are commonly missing in the radio bands. 
For instance, although a jet-like region is found in the radio Crab PWN, this structure bifurcates from a spot of the X-ray jet; for Crab, Boomerang and B1706 PWN, radio torii are found enveloping the X-ray torus/jet structures in the outer region \citep{2017ApJ...840...82D,2006ApJ...638..225K,2023ApJ...945...82L}; radio wisp-like structures inside 3C~58 are also seen which rarely coincide with X-ray structures \citep[e.g., wisps and loops; see][]{2006ApJ...645.1180B}.

Magnetic fields are an important factor of particle transportation in PWN.
Radio polarization observations are a reliable method to analyze PWN $B$ 
-field configurations. For instance, a toroidal $B$-field configuration is commonly seen in the equatorial regions of some middle-aged PWN \citep[e.g., Vela, Boomerang, B1706 PWN; ][]{2003MNRAS.343..116D,2006ApJ...638..225K,2023ApJ...945...82L}.  
However, we note again that PWN structures in radio observations are often misaligned with the X-ray PWN, and the $B$-field of a PWN jet is still mysterious as a jet is rarely detected in radio bands. 
Recently, \emph{IXPE} polarization observations allowed the inference of the $B$-field morphology of the Crab and Vela Nebulae in the X-ray band \citep{2022Natur.612..658X,2023NatAs...7..602B}, but the results show limited capability to analyze smaller scale structures.

In this paper, we present radio and X-ray studies of the PWN-SNR system \gname. 
The \gname\ PWN is powered by the hard X-ray/ soft gamma-ray pulsar  PSR J1811$-$1925, located near the center of a {4\arcmin} supernova remnant shell. 
Timing observations with \emph{the Advanced Satellite for Cosmology and Astrophysics (ASCA)} measured an X-ray pulsar period of 65\,ms and a spin-down luminosity of $\sim8.8\times10^{36}$\,ergs\,s$^{-1}$ \citep{1999ApJ...523L..69T}, 
while in the radio band, a 5.5 hour Green Bank Telescope observation at S band using the Guppi back end in 2009 failed to detect the pulsar. 
Expansion of the SNR shell implies that this system is around 2000 years old, making it possibly the third or fourth youngest galactic pulsar \citep{2003ApJ...588..992R,2016ApJ...819..160B}.
In radio bands, an overall east-west elongated PWN was found near the SNR center with a possible torus/jet radio feature found close to the X-ray pulsar region, and complex filaments are seen further away \citep{2003ApJ...588..992R}.  \emph{Chandra} observations also show  torus + jets features close to the pulsar (Figure \ref{fig:jet_torus_g11}). Relativistic motions of bright spots and an apparent Doppler boosting effects were seen along the X-ray PWN elongation, indicating that the bright bar close to the pulsar should be the jet structure. The Mg K$\alpha$ line map from more recent \emph{Chandra} observations implies that the central PWN has already interacted with the reverse shock \citep{2016ApJ...819..160B}.

Here we revisit the radio PWN \gname\ with new Australia Telescope Compact Array (ATCA) radio imaging and polarization observations and combine them with archival X-ray observations to study the broad band spectrum of the PWN, focusing on the small scale structure, to further understand the origins and behaviors of the high energy particles inside.   
Section \ref{sec:data} discusses the radio and X-ray observations and data reduction performed for this study. We show our main results in Section \ref{sec:result} and discuss them in Section \ref{sec:discussion}. 
All these are finally summarized in Section \ref{sec:conclusion}.  
\begin{table*}[ht!]
    \centering
    \begin{tabular}{cccccc}
    \toprule
        Obs.   & Array  & Center Freq. & Bandwidth & No. of   & Integration  \\
          Date         & Config.& (MHz)        & (MHz)     & Channels & Time (hr)   \\
                  \hline
        3\cm  &  &  &  &  &  \\ \hline
         2022 Oct 4& 6D & 5500  & 2048 & 2049  & 10.8 \\ 
         2022 Nov 25& 6C & 5500 & 2048 & 2049 & 9.9 \\ \hline
         6\cm  &  &  &  &  &  \\ \hline 
         2022 Oct 4& 6D & 9000 & 2048 & 2049 & 10.8 \\
         2022 Nov 25& 6C & 9000 & 2048 & 2049 & 9.9 \\
         \hline
         16\cm  &  &  &  &  &  \\ \hline
         2024 Nov 17 & 6A & 2100 & 2048 & 2049 & 10.0 \\
         \hline
    \end{tabular}
    \caption{ATCA Observations of \gname\ in this study}
    \label{tab:obs}
\end{table*}
\section{Observations and Data Reduction}
\label{sec:data}

\subsection{Radio Observations}
We performed new radio observations of G11.2$-$0.3 PWN with ATCA on 2022 October 4 and November 25 as project C3503, as well as on 2024 November 18 as project C3644.
All observations used 6\,km array configurations of ATCA (6A, 6C, and 6D) to prioritize good high resolution \uv\ coverage to compare with \emph{Chandra} X-ray observations. 
The C3503 observations took data at 3 and 6\cm\ bands simultaneously with frequency centers at 5500 and 9000\,MHz, and C3644 observations took 16\,cm data.
As the observations were carried out after the CABB upgrade \citep{2011MNRAS.416..832W}, observations in each band have a frequency bandwidth of 2048\,MHz enabling good sensitivity and also improved coverage in the \uv\ space.
The two observations together cover a \uv\ distance of 0.6-185\,k$\lambda$ at 3\cm\ and 0.2-125\,k$\lambda$ at 6\cm; for the 16\,cm observation, it covers a \uv\ distance of 0.5-60\,k$\lambda$. These observations enable good imaging at scales of a few arcseconds but have relatively sparse baseline coverage for imaging at scales of a few arcminutes.   
The 3, 6, and 16\,cm\ data have total integration times of 20.7, 20.7, and 10.0\,hr to reach good data qualities. 
1934$-$638 was used as the primary calibrator and 1829$-$207 as the secondary calibrator for the 3 and 6\,cm observations, while 1830$-$210 was used as the phase calibrator for the 16\,cm observation. 
The observation parameters are listed in Table \ref{tab:obs}.

The MIRIAD is a package to reduce, analyze and plot radio intereferometric data originally developed for use with ATCA \citep{1995ASPC...77..433S}. We use it for most of the radio analyses in this study. 
Each raw dataset had 40 channels flagged on the edges of the observing band, and we also flagged channels with severe radio frequency interference (RFI) before calibration. 
We used standard procedures within MIRIAD to obtain bandpass, gain, and flux scale solutions from the calibrators. Further RFI flagging was done during calibration as warranted.

The same, single pointing center was used  for all the PWN observations (the central pulsar), and  the whole SNR region was contained within the primary beam at all frequencies. 
For the 3 and 6\,cm data, after calibration solutions were  applied to the target data, we combined the two epochs of observations and generated Stokes \textit{I} dirty images of the \gname\ PWN for each frequency band. 
A maximunm entropy routine was used to clean the dirty images. Multi-frequency synthesis was used for data imaging in each band and we used a weighting that was inversely proportion to the noise level. 
More detail about the radio imaging is in Section \ref{sec:result}. 

\begin{table}[h!]
    \centering
    \begin{tabular}{cccc}
    \toprule
    Date                               & Observation ID & Effective  \\ 
                                       &                &      exposure time (ks)               \\ \hline
    2013 May 05-07                     & 14831                  & 173                         \\ 
    2013 May 25-26                     & 14830                  & 58                          \\ 
    2013 May 26-27                     & 14832                   & 63                             \\ \hline
    \end{tabular}
    \caption{Chandra Observations of \gname\ in 2013}
    \label{table_x-ray}
\end{table}

\subsection{X-ray Observations}

We also chose three archival \emph{Chandra} X-ray observations  from 2013 to perform a multi-band analysis with our radio observations. 
Detailed information of the observations is stated in Table \ref{table_x-ray}. 
All three observations were performed with the ACIS S3 CCD chip in May 2013, and the total exposure time was 294\,ks. We merged all  of these observations to match the reference frame of the longest pointing (OBS. ID 14831). CIAO v4.16 and CALDB~v4.11 were utilized to process all the X-ray data.  Results are described in the next section.

\section{Results}
\label{sec:result}
\subsection{Morphology}

\begin{figure*}
    \centering
    \includegraphics[width=1\linewidth]{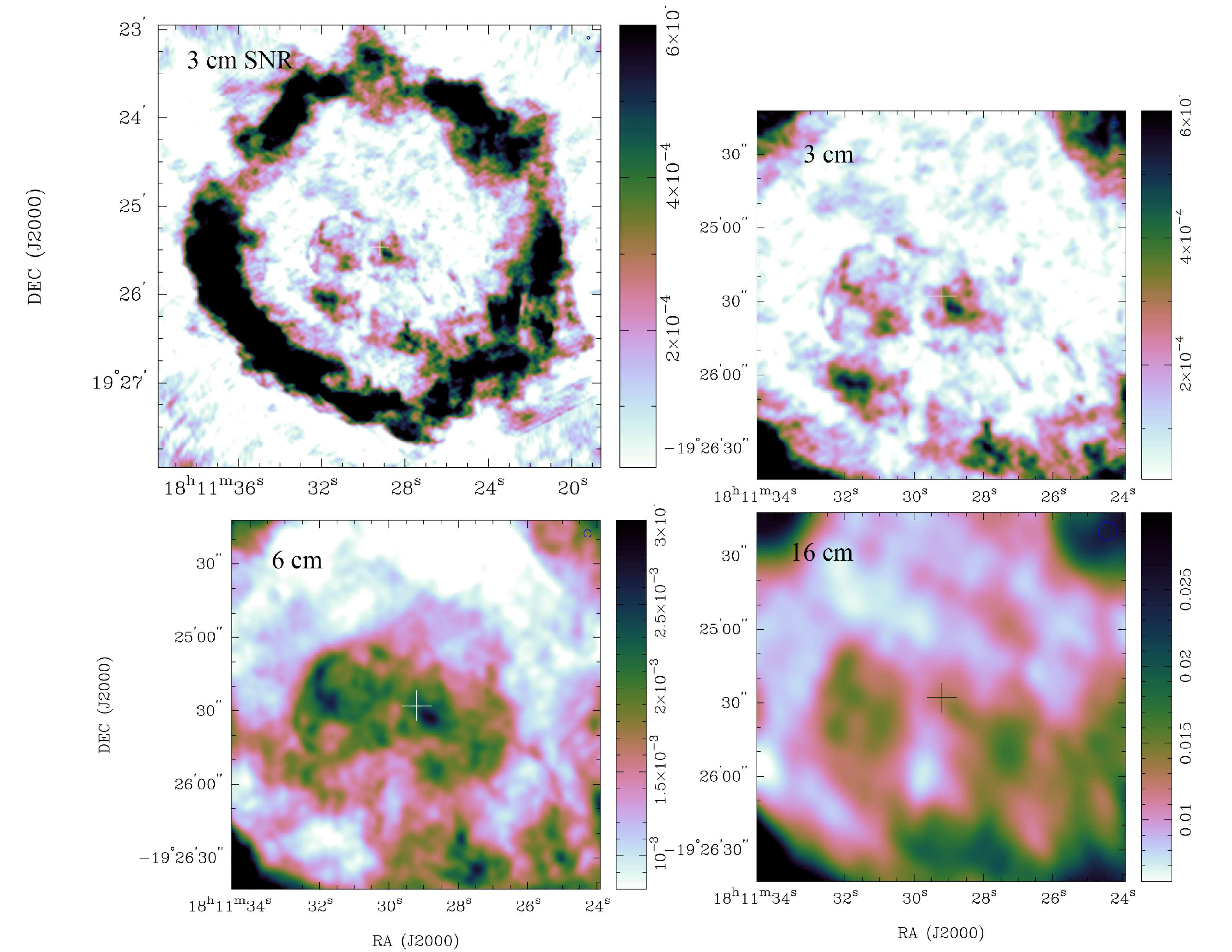}
    
    \caption{ATCA images of the \gname\ region. The \textit{top left} panel covers the whole 3\,cm SNR, and the other panels show the zoomed-in images of the PWN at 3\,cm (\textit{top right}), 6\,cm (\textit{bottom left}), and 16\,cm (\textit{bottom right}), respectively. The white crosses indicate the position of PSR J1811-1925; the circles in the corners show the beam size in the maps; the color bars on the right are in units of Jy\,beam$^{-1}$.}
    \label{fig:3_6cmI}
\end{figure*}

\begin{figure*}[ht!]
    \centering
    \includegraphics[width=0.6\linewidth]{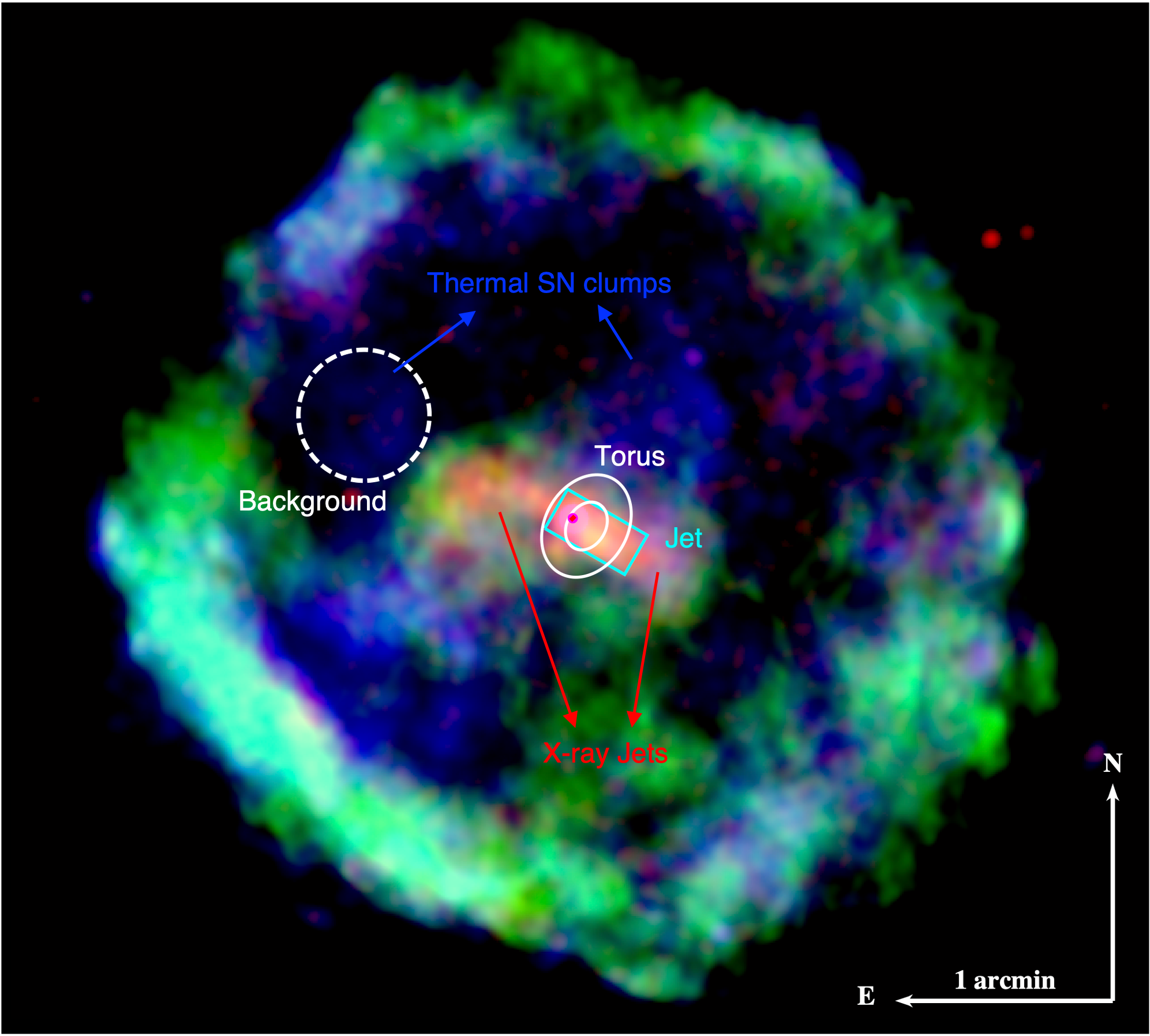}
    \caption{RGB image of \gname\ SNR comparing 6\,cm ATCA radio map (\textit{green}), 2.7-9.0\,keV \emph{Chandra} X-ray map (\textit{red}), and 0.5-2.0\,keV \emph{Chandra} X-ray map (\textit{blue}) with smoothed FWHMs of around 5\arcsec. 
    The magenta circle, cyan box, and white elliptical annulus indicate the central pulsar, radio jet and torus, respectively. 
    The white dashed circle is used as a background region for flux measurements of PWN structures.    }
    \label{fig:radioX}
\end{figure*}

Figure \ref{fig:3_6cmI} shows the total intensity maps of the \gname\ region at 3, 6, and 16\,cm. 
As is mentioned above, data in all bands are weighted inversely proportional to visibility noise and multi-frequency synthesis was performed; the dirty maps were generated with task \texttt{invert}, and  the cleaned images were generated with \texttt{mosmem} and \texttt{restor}. 
For 3 and 6\,cm observations, we fix robust parameters at 0.0 and FWHMs of 2\arcsec\ and 3\arcsec\ to optimize both visibility and resolution \citep{1995AAS...18711202B}; the 16\,cm data is plotted with a robust of -0.5 and FWHM of 7.5\arcsec; after cleaning, the models are restored to produce cleaned images with the same FWHMs, and the results experience primary beam corrections with \texttt{puthd} and \texttt{linmos}. 
The cleaned images have rms noises of 0.03, 0.08, and 0.52\mJyPerBeam\ at 3, 6, and 16\,cm, respectively.
The outer rim of the \gname\ SNR generally has a circular geometry with a diameter of $\sim250\arcsec$ (see Figure \ref{fig:3_6cmI} \textit{top left}), and a shell thickness of $\sim$20\arcsec , with the south-east (SE) part being brighter and denser than the rest of the shell.
Both the pulsar and the radio PWN are located near the center of the SNR. The PWN is east-west elongated with a size of $\sim 1.5\arcmin\times0.7\arcmin$. Several larger scale SNR structures within the shell are also detected. For example, the filamentary structure with a thickness of 20\arcsec\ in the SE part of the SNR. 
In the SNR center, all observation bands show a similar radio PWN morphology close to the pulsar.
A linear structure (most significantly at 6\,cm) extends around 0.5\arcmin\ from the pulsar position to the southwest (SW), along which a bright spot close to the pulsar is also detected in both 3 and 6\cm\ bands, while fainter emissions are detected on the opposite direction (NE) of the pulsar. 
The 6\cm\ image also resolved a compact torus-like structure enveloping the pulsar position, with a size of $\sim20\arcsec\times10\arcsec$ perpendicular to the linear structure elongation, similar to what is discovered in previous VLA observations \citep{2003ApJ...588..992R}. 
A similar structure is resolved in the 3\cm\ observations, whereas not at 16\cm\, probably due to insufficient resolution.  
The new observations also detect two arc-like structures with a radius of $\sim20\arcsec$ detected around $0.5\arcmin$ northeast from the pulsar in all bands consistent with the arc-like structures detected in the VLA observations.

Figure \ref{fig:radioX} compares the \emph{Chandra} X-ray observations of the \gname\ SNR with the recent ATCA radio observations. 
Previous X-ray studies indicated that the thermal spectrum of the PWN region becomes negligible around 3\,keV \citep{2016ApJ...819..160B}, which is also confirmed in Section \ref{subsec:res_spec} below. 
Here, for the \emph{Chandra} X-ray observations, we plot 2.7-9\,keV data in red and 0.5-2.0\,keV data in blue to show the different components; then we plot the 6\,cm ATCA image in green for comparison. 
The non-thermal X-ray torus/jet and pulsar features are resolved as red components in the center of SNR; it should be noted that the NE X-ray jet turns to SE around 0.5\arcmin\ from the pulsar when it reaches the radio arc. Non-thermal emission is also found throughout the SNR shell but at a much lower surface brightness than  in the central PWN. 
The radio PWNs show spatial coincidence with the X-ray torus/jet feature, especially for the SW jet and the spot inside; however, no significant spatial connection is discovered in the  outer parts of the PWN. However, filamentary radio structures are detected along the outline of the X-ray torus/jet structure (e.g., the arc-like features on the east part).
Intricate large-scale structures inside SNR are detected in both radio and X-rays. A $\sim0.5\arcmin$ clump at 0.5-2\,keV is located just beyond the X-ray NE jet and radio arc. In the 0.5-2\,keV map, we also find another diffuse structure at the NE end of the SE radio filament (mentioned in Figure \ref{fig:3_6cmI}), as well as a semi-torus-like feature NW of the PWN. 
Both radio and X-ray observations show the SNR rim, and these structures generally align, although with some interesting discrepancies. For example, a bubble feature at the north edge if the shell is seen as a circle in the radio images, but only the Southern half of the bubble is prominent in the X-ray emission.

\subsection{Spectrum}
\label{subsec:res_spec}

\begin{figure}[h!]
    \centering
    \includegraphics[width=0.9\linewidth]{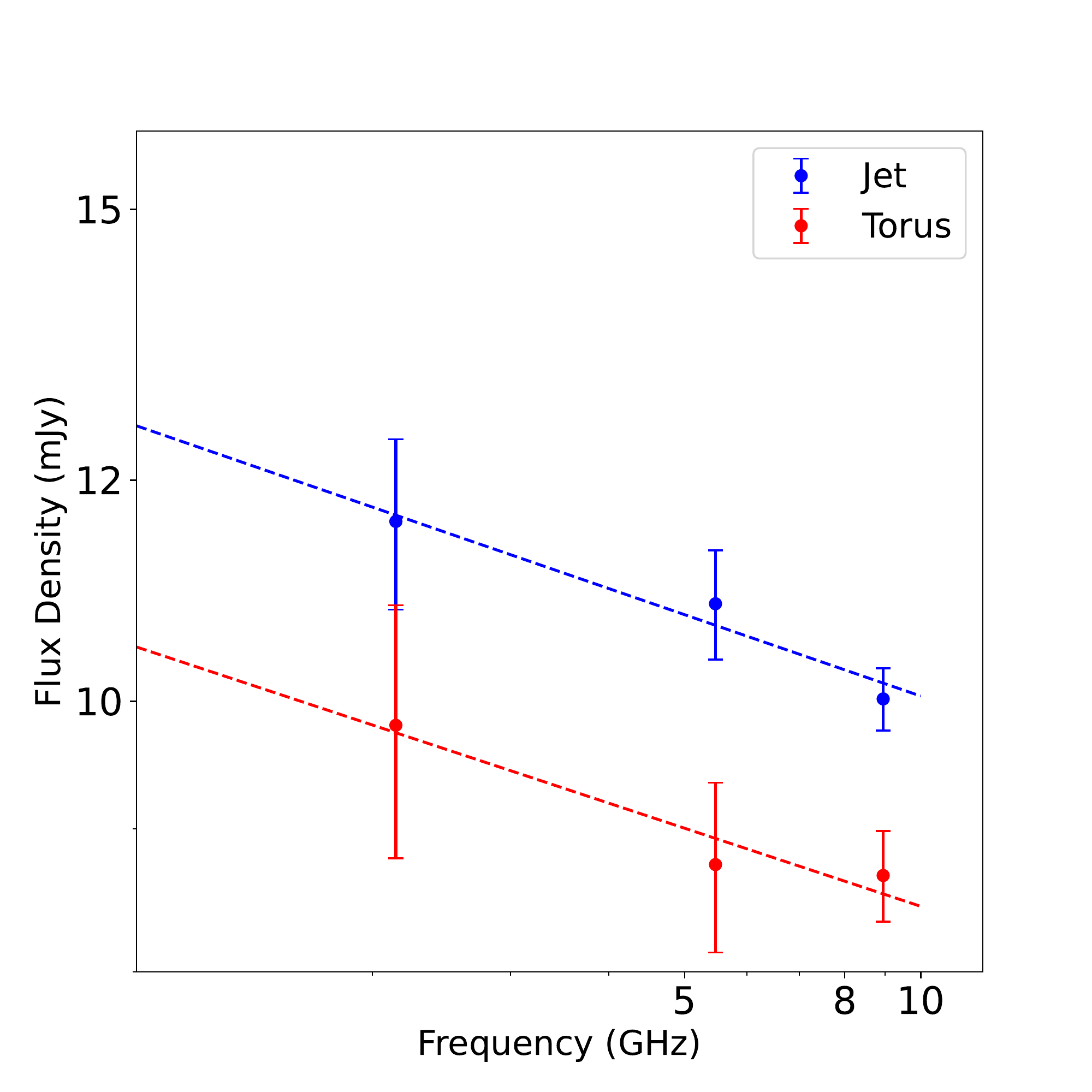}
    \caption{Radio spectra of the Jet and Torus region of the PWN. }
    \label{fig:radio_spec}
\end{figure}

High resolution radio observations are helpful to resolve the counterparts of the X-ray PWN structures in detail. 
We therefore measured both radio and X-ray flux densities of the PWN  features (e.g., torus, jet).
As the new ATCA observation have sparse coverage at short $u-v$ distances, there would be significant missing flux problems for large scale structures (like the SNR), especially as compared to the previous VLA images. However, there should be very little missing flux at the scale of the inner PWN. 
Since there is enhanced radio emission throughout the shell,  we select a nearby filament free region within the SNR shell (see in Figure \ref{fig:radioX}) as a background region for subtracting the SNR emission from the PWN emission.

\subsubsection{Radio Spectrum}
Figure \ref{fig:radio_spec} shows radio spectra of the different \gname\ inner PWN structures from our observations. 
We select the regions indicated in the left panel of Figure \ref{fig:radioX} to measure the torus and the SW jet. 
As the radio pulsar has not been detected even with long observations with the Green Bank Telescope, we do not need to measure and exclude contamination from pulsar emission.
We measure flux densities of the jet region of $10.0\pm0.3$,  $10.8\pm0.5$, and $11.5\pm0.8$\,mJy at 3, 6, and 16\,cm respectively; and of $8.7\pm0.3$, $8.7\pm0.6$, $9.8\pm1.0$\,mJy for the 3, 6, and 16\,cm torus. 
This corresponds to flux spectral indices $\alpha$ (where $F\propto \nu^{\alpha}$) for the jet and torus of $-0.10\pm0.01$ and $-0.09\pm0.01$. This is somewhat flatter than the $\alpha \sim -0.25$ for the overall PWN measured with the VLA by \citet{2002ApJ...572..202T}. Examination of their spectral tomography maps shows indications of a flatter spectrum in the torus/jet region, consistent with the ATCA measurements. 
Future measurements of this source at different frequencies would further clarify the spectral properties of the various radio features seen in the PWN.  

\subsubsection{X-ray Spectrum}

\begin{table}[h!]
    \centering 
    \footnotesize
    \begin{tabular}{ccccc} 
    \toprule
    Region & $\Gamma$ & Amplitude \footnote{ in a unit of (10$^{-5}$)} &N$_H$\footnote{ in a unit of (10$^{22}$\,cm$^{-3}$)}
    &flux \footnote{ unabsorbed flux; in a unit of (10$^{-13}$\,erg\,cm$^{-2}$\,s$^{-1}$)} \\
    \midrule
    \multicolumn{2}{l}{0.9-7.5\,keV} &  & & \\\midrule
    Jet    & $1.70\pm0.05$  & $9.86\pm0.64$& $2.62\pm0.12$   & 4.51$^{+0.03}_{-0.04}$ \\
    \hline \multicolumn{2}{l}{4-7.5\,keV} \\ \midrule
   Jet  & $1.73\pm0.14 $ & $9.83\pm2.29$ & 2.62(frozen)   & 1.58$^{+0.37}_{-0.48}$\\
   Torus    & $1.78\pm0.25$   & $2.96\pm1.17$ & 2.62(frozen)   & 0.44$^{+0.16}_{-0.24}$ \\
    \bottomrule
    \end{tabular}
    \caption{Fitting of non-thermal components of X-ray PWN structures}
    \label{table_xresult}
\end{table}
\begin{figure}[ht!]
    \centering
    \includegraphics[width=1\linewidth]{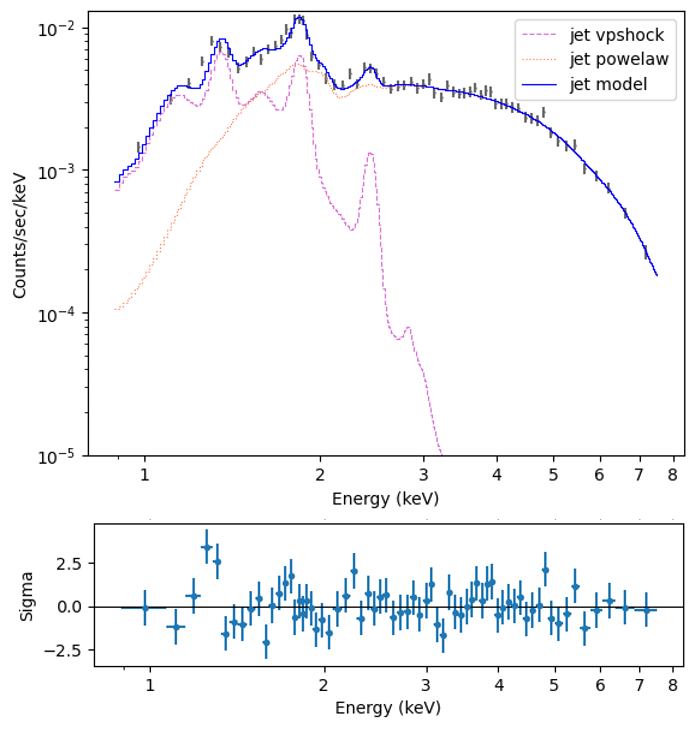}
    \includegraphics[width=1\linewidth]{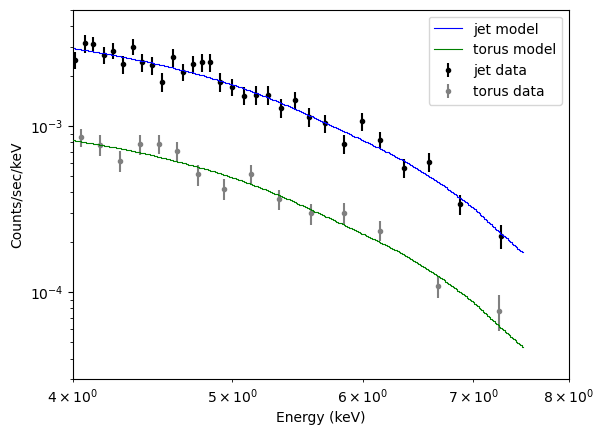}
    \caption{\emph{Chandra} X-ray spectra of PWN structures as indicated in Figure~\ref{fig:radioX}. The top panel shows the 0.9-7.5\,keV spectrum of the X-ray SW jet and fitting results attached with a $\sigma$ residual spectrum. The blue, pink, and orange curves represent the total, vpshock, and power-law models, respectively. The bottom panel shows 4-7.5\,keV SW jet and torus spectra fit with a non-thermal model, the blue and green curves indicate to models of jet and torus, respectively.  }
    \label{fig:xspectrum}
\end{figure}

We also analyzed the spectra of the X-ray features coincident with the radio features near the pulsar.  The selected observations are of sufficient exposure time for detailed spectral spectral studies of compact regions. 
Lines in the spectra indicate that the large-scale thermal radiation related to the shell and filaments of the SNR is detected within in jet area \citep{2016ApJ...819..160B}.   
We therefore  extract the spectrum of a region with only large-scale SNR ejecta inside to characterize this thermal emission. 
The result of fits to this region shows a dominating thermal component only at energies below 3\,keV, and this is applied to the analysis below.

We then study the spectra in the torus and jet regions with spectral fitting; the source region are chosen as shown in Figure \ref{fig:radioX}. 
We used task \texttt{specextract} to extract spectra in these regions, and a source-free region in the SNR as background region, as mentioned above. 
The overlapping regions are excluded to prevent possible contamination in spectral analyses.
Since we are primarily interested in the power law component of the spectra, which is presumably dominated by the PWN, we analyze the spectra in the 0.9-7.5\, keV range. At higher energies, the photon count is too low to further constrain the spectra, and at lower energies,  X-ray emission is heavily absorbed and overly affected by spectral lines dependent on individual element abundances. 
As the thermal component is significant up to several keV, we first choose a model of \texttt{xsphabs}$\times$(\texttt{vpshock+powlaw1d}) to fit the jet spectrum, which is shown in the top panel of Figure \ref{fig:xspectrum}. 
The spectrum clearly shows that the energy of thermal radiation is mainly below 4\,keV, while the spectrum above 4\,keV is dominated by non-thermal emissions. 
The fit result for the non-thermal emission from the jet is in Table \ref{table_xresult} for reference.

We then simply take 4-7.5\,keV data as non-thermal emission for subsequent fitting with the model of \texttt{xsphabs}$\times$\texttt{powlaw1d}, freezing the 
$N_\mathrm{H}$ value at 
2.62$\times10^{22}$\,cm$^{-3}$ 
from the jet fitting described above.
The bottom panel of Figure \ref{fig:xspectrum} shows the X-ray spectra and fitting in the jet and torus regions.
Our result of $\Gamma = 1.73$ in the jet, matches the fit result of 1.75 in previous studies \citep{2016ApJ...819..160B}, while we note that our regions are much closer to the pulsar.
Compared with the radio spectrum, the X-ray spectrum is expected to intersect with the radio spectrum, and more information about the multi-frequency spectrum is discussed in Section \ref{sec:discussion}.
\begin{figure*}[ht!]
    \centering
    \includegraphics[width=1\linewidth]{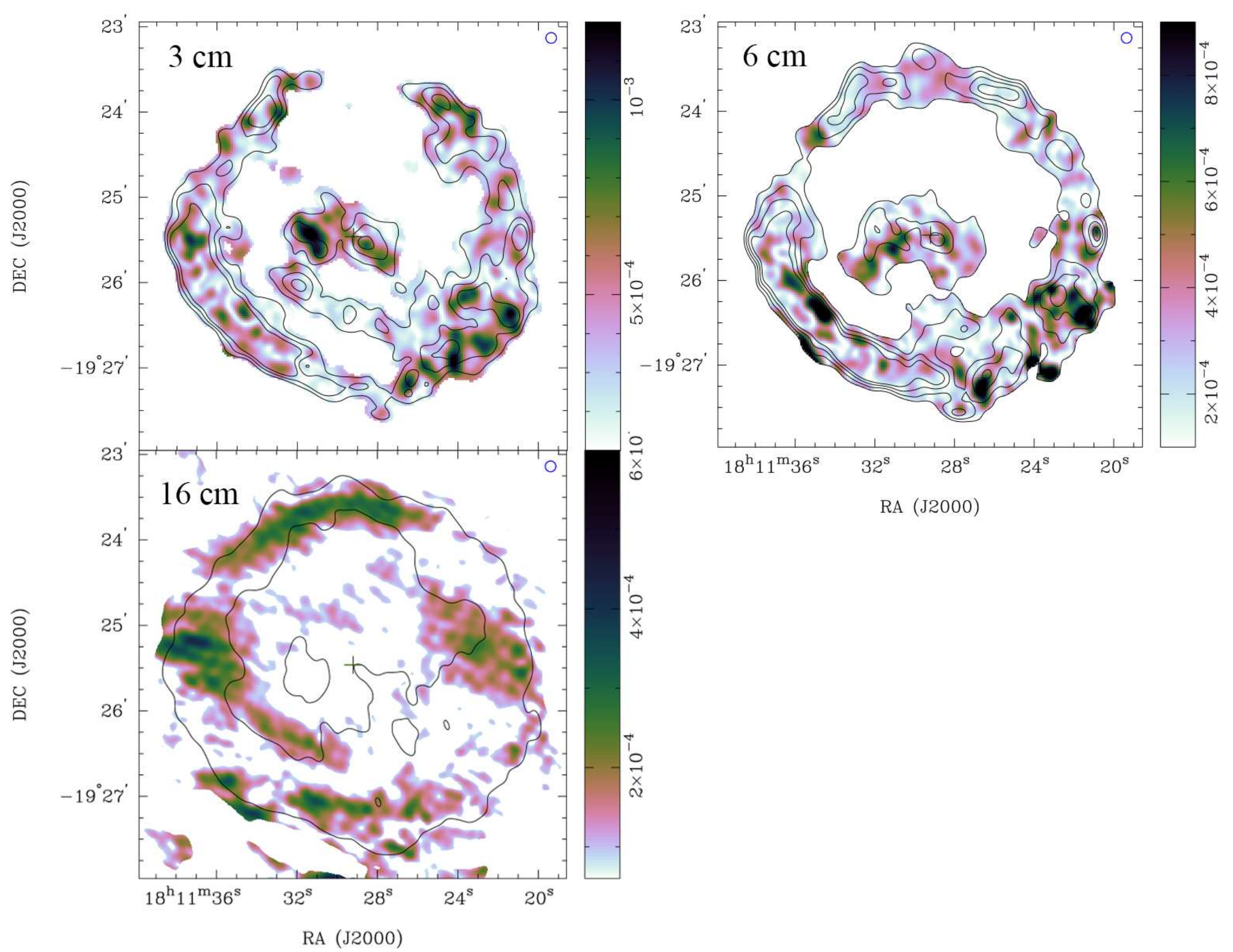}
    \caption{PL images of the SNR \gname\ at 3\,cm (\textit{upper left}), 6\,cm (\textit{upper right}), and 16\,cm (\textit{lower left}), with color bars on the right in units of \JyPerBeam. The crosses in the center indicates to the pulsar position, and the contours are total intensity of 4.2, 6.0, and 8.0\mJyPerBeam\ for 3\,cm; 7.5, 10.0, 12.5 and 15.0\mJyPerBeam\ for 6\,cm; and at 13\mJyPerBeam\ for 16\,cm.  }
    \label{fig:3_6cm_PI}
\end{figure*}

\subsection{Polarimetry}

Figure \ref{fig:3_6cm_PI} shows the polarization intensity (PI) images of \gname\ at 3, 6, and 16\,cm, we used task \texttt{invert} to generate Stokes \textit{I, Q, U} dirty maps at all three wavelengths, with all having FWHMs of 7.5\arcsec\ for consistent analyses and to have higher signal to noise per beam. We then generate cleaned Stokes \textit{I, Q, U} model maps with task \texttt{pmosmem} and then plot cleaned images with the same FWHM. 
For both 3 and 6\,cm observations, rms noises of Stokes \textit{I} maps are around 
0.25\mJyPerBeam
, and for Stokes \textit{Q, U} maps around 
0.03\mJyPerBeam. We plot PL images with \texttt{impol}, clipping polarized intensities below $3\sigma$ at 3 and 6\,cm and regions that do not have a high S/N in the total intensity maps ($\sigma>25$ for 3\,cm and $\sigma>30$ for 6\,cm) to better identify regions of significant PI. For the  16\,cm maps, the rms of the total intensity map and PL map are 0.6 and 0.03\mJyPerBeam, and we clip at $3\sigma$ for the Stokes $I, Q, U$ maps.  
The PWN is not very highly polarized, with average polarization fractions of $\sim 10\%$ and $\sim 3.5\%$ for 3 and 6\,cm observations.
At 3\,cm, the PI map roughly follows the torus/jet morphology of the PWN, with a $\sim1\arcmin$ NE-SW linear feature going through the pulsar region in the direction of the jets and spots close to the torus region; the polarization emission close to the pulsar peaks in the SW jet region; and extended PL emission is also significant on both sides of the NE jet (i.e., the arc).
For the 6\,cm observations, PL emission is also found in the SW jet region and close to the pulsar; there are some extended PL features southeast of the pulsar, while some fainter features are on the other side (NW of the pulsar).
The 16\,cm\ observation also shows polarized emission in the SNR shell. 
However, no PL emission is detected in the PWN region.
We note that this could be due to heavy depolarization effects at 16\,cm, because of the wider wavelength coverage in this observation. 
We illustrate the impact of such an effect below in the paragraph about rotation measure in PWN.
Although beam depolarization could be an alternative cause, the smoothed 3 and 6\,cm PL maps (to a FWHM of 7.5\arcsec) still show the PWN geometry having significant intensities, precluding this scenario.

\begin{figure*}[ht!]
    
    \centering
    \includegraphics[width=1\linewidth]{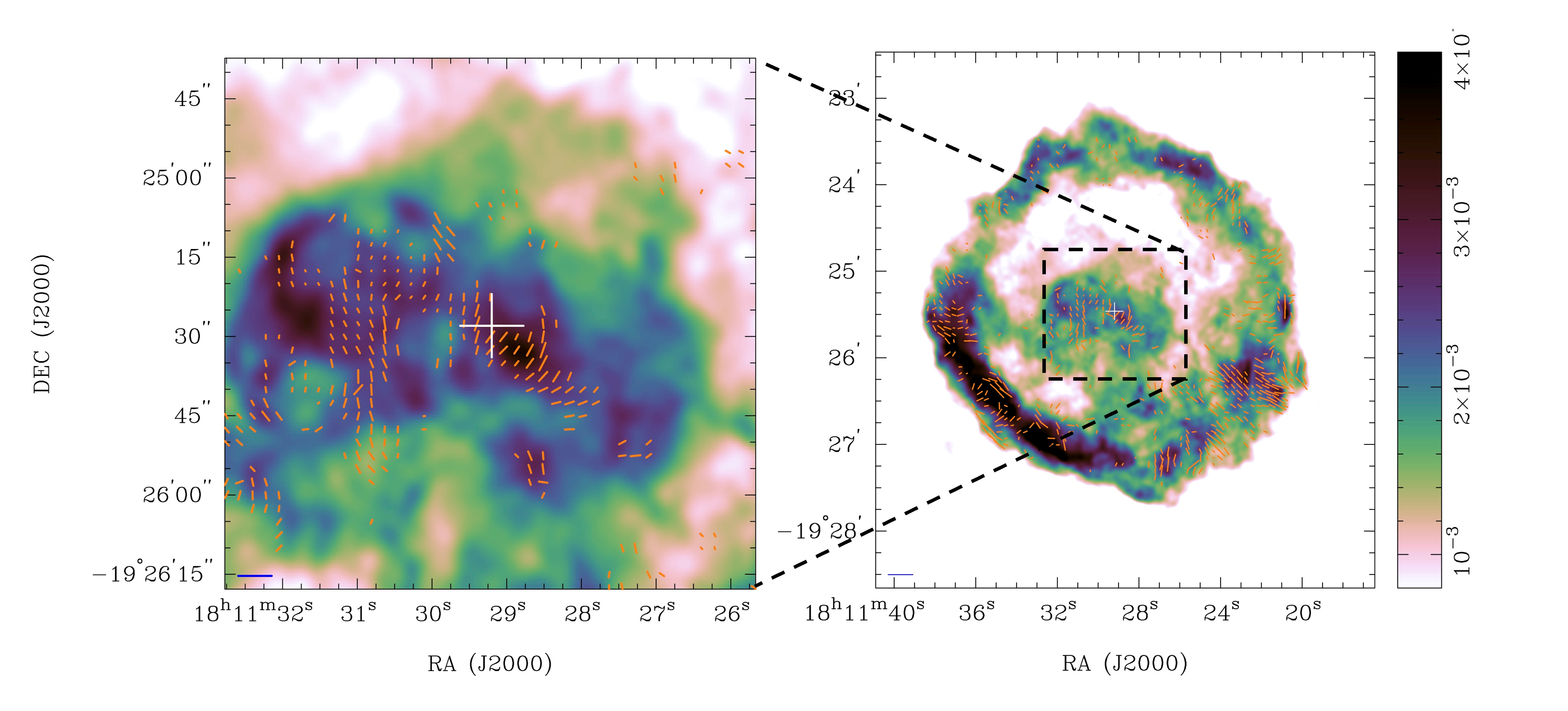}
    \caption{Image of magnetic field configuration of \gname\ SNR region obtained from ATCA observations. The right panel shows the ATCA 6\,cm Stokes \textit{I} map of the SNR, and the left panel zooms in to reveal the PWN region. The vectors indicates to the $B$-field directions in the map, the length of which reveal PL intensities with a bar on the bottom-left of 0.5\mJyPerBeam\ as a reference. }
    \label{fig:B-field_dir}
\end{figure*}

Figure \ref{fig:B-field_dir} reveals for the first time the $B$-field configuration of this PWN in detail.
We exclude the 16\,cm PL results in the fitting of the $B$-field due to possible depolarization problems mentioned above.
Instead, we split the 6\,cm observation by frequency for, with the frequency center of the split groups being 4988 and 6012\,MHz, to increase the number of frequency bands with optimized visibilities.  These data are included for fitting together with the 3\,cm PL data. 
We obtain PL images with FWHM of 5\arcsec\ in all wavelengths to optimize for resolution and sensitivity, and use \texttt{imrm} to perform a linear fitting of the Faraday rotation effect: 
\begin{equation}
    \chi_{\lambda}=\chi_0+RM\cdot\lambda^2, 
    \label{eq:faraday}
\end{equation}
where $\lambda$, $\chi_\lambda$, and $\chi_0$ refers to emission wavelength, observed PL direction, and the original PL direction, respectively. $RM$ refers to the Faraday rotation measure (RM) of the interstellar medium.
At large scales, the SNR shell generally shows a radial $B$-field (especially in the SW part) after correcting the Faraday rotation component, while the $B$-field directions are a bit turbulent in the SE part (the peak of SNR rim). 
Such a configuration agrees with the previous low resolution result obtained with the Effelsberg 100-m radio telescope \citep{2001A&A...372..627K}. 
Zooming in to the PWN region (Figure \ref{fig:B-field_dir} left), the $B$-fields in the jet regions is observed to be perpendicular to the jet propagation direction (i.e. a helical $B$-field). The $B$-field in the torus region is not clearly revealed in these observations, which could possibly be due to significant PL depolarization when the $B$-field geometry projecting along line of sight is complex (i.e. beam depolarization). 
The outer regions show more complicated $B$-field geometries; interestingly, the $B$-field on the both sides of the NE jet tend to be parallel to the jet elongation.

\begin{figure}[h!]
    \centering
    \includegraphics[width=1\linewidth]{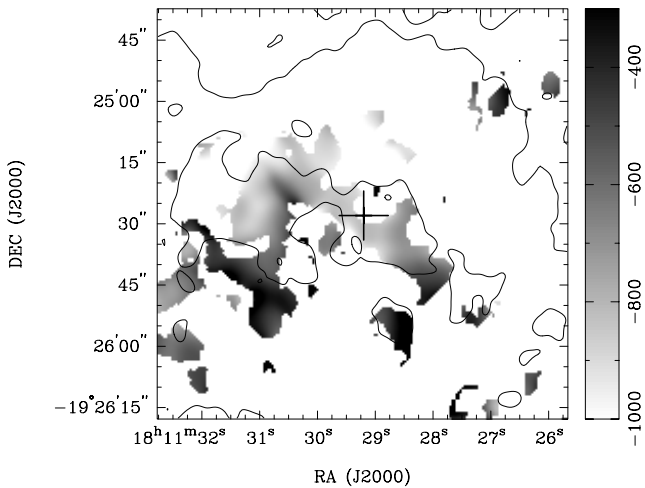}
    \caption{The map of Faraday rotation measure in the PWN region. The contours and cross are as introduced above, and the gray-scale color bar shows the RM values with a unit of rad\,m$^{-2}$. }
    \label{fig:rm}
\end{figure}

Figure \ref{fig:rm} shows RM distributions of the PWN region, RM values within the PWN range from $-1000$ to around $-200$\,rad\,m$^{-2}$ and the average value is $\sim-700$\,rad\,m$^{-2}$.
Along the jet region, RM decreases smoothly from $-400$\,rad\,m$^{-2}$ at the SW end down to around $-900$\,rad\,m$^{-2}$ in the NE jet region, and RM close to the pulsar region is $-830$\,rad\,m$^{-2}$, and this result agrees with previous low resolution Effelsberg PL observation of this source.
Given such an RM, we illustrate that the 16\,cm observation could have a half angle of around 42\,rad in Appendix \ref{sec:depol}, and a depolarization factor of 99\%; even if we split frequency bins to 20\,MHz width, this confirms the suggestion of heavy bandwidth depolarization in this band; the depolarization is still as high as around 70\%, when the sensitivity in one single bin has become very low.
The RM map also shows that RM values in other structures close to the pulsar fluctuate smoothly, with an average of -670\,rad\,m$^{-2}$.
The outer shell of the SNR also shows variation between $\sim-1000$ and $\sim200$\,rad\,m$^{-2}$, and it has a average RM value of $-500$\,rad\,m$^{-2}$.

\section{Discussions}
\label{sec:discussion}

\subsection{Magnetic fields in \gname\ PWN}

This study revealed a radio jet feature in the  \gname\ PWN with a helical magnetic field inside.  Before this,  no similar source had been observed with such a feature.
The lack of observed radio jets in PWNe, and hence polarization studies, has hindered work on magnetic features in PWN jets observed in X-rays (e.g., in numerical simulations). 
This has limited discussion focused on the morphology and $B$-field configuration of PWN jets. In most numerical models, some of the higher latitude plasma in the toroidal outflow is channeled to the polar regions by the magnetic hoop force of the local toroidal $B$-field, and then forms a collimated jet \citep[e.g., the jet of Crab Nebula;][]{2003MNRAS.344L..93K,2014MNRAS.438..278P}. It is therefore expected PWN jets would have a helical $B$-field structure. 
Previous radio observations of DA~495 revealed a double-lobed radio PWN. Polarization measurements suggest that this system has a poloidal $B$-field along the putative pulsar spin axis, in addition to a typical toroidal $B$-field, \citep{2008ApJ...687..516K}; however, no significant jet feature was detected in the system. 
While further discussion about the helical $B$-field in the PWN jet goes beyond the scope of this study, we note that a $B$-field geometry is commonly detected in AGN jets \citep{2002PASJ...54L..39A,2023NatAs...7.1359F}, where the rotation of plasma in the vicinity of the central black hole or accretion disk could form a helical $B$-field. We hope this study may inspire further detailed models of local $B$-field features in PWN.

We also estimate the magnetic field strength close to the jet region with a minimum $B$-field energy assumption \citep{2004IJMPD..13.1549G}. 
In this method, the equipartition magnetic field strength $B_{eq}$ is estimated as: 
\begin{equation}
    B_{eq} = [6\pi(1+k)c_{12}L_{syn}\Phi^{-1}V_{PWN}^{-1}]^{2/7},
\end{equation}
where $V_{PWN}$ is the emission volume, $L_{syn}$ is the synchrotron luminosity, $\Phi$ is a filling factor for the emission (it is usually taken as 1 even though not 100\% of the volume emits), $k$ is the ratio between the electron energy and the energy of heavy particles, and $c_{12}$ is a constant related to synchrotron radiation process and weakly depends on the frequency range \citep{1970ranp.book.....P}. 
We take the radio jet spectrum until the break near 10$^{11}$\,Hz (also see in Section \ref{subsec:multi-freq Spec}) to be the main channel of SYN energy loss, as is estimated above.
Given a distance of $\sim$5\,kpc \citep{1985ApJ...296..461B}, this would suggest a luminosity $L_{syn}$ of 
2.7$\times$10$^{31}$\,erg\,s$^{-1}$ 
integrating from 10$^7$ to $10^{11}$\,Hz. Modeling the radio jet as a $15\arcsec\times8\arcsec$ cylinder, the volume of the jet would be 6.16$\times$10$^{53}$\,cm$^3$.
If we assume equipartition in this region, the above values suggest
\[B_{eq}\simeq85(1+k)^{2/7}\Phi^{-2/7}d_5^{-2/7}\,\mu\mbox{G},\]
where $d_5$ is in a unit of 5\,kpc. This is of the same order as the typical value ($\sim10\,\mu\mbox{G}$) among PWNe \citep{2016ApJ...820..100M,2023ApJ...945...82L}. It is unsurprising that we detect a somewhat stronger $B$-field within the compact jet region close to the pulsar.

\subsection{Multi-frequency Spectral Analysis} \label{subsec:multi-freq Spec}

As is mentioned in Section \ref{sec:intro}, particle acceleration mechanisms in PWNe are still intensively debated. Two leading theories are Fermi acceleration and magnetic reconnection. 
Particles experiencing Fermi acceleration are believed to initially emit a simple power law synchrotron spectrum up to some maximum energy. After some time period, a continuously operating Fermi process would result in broken power-law with a clearly defined break due to the higher energy particles cooling faster than lower energy particles.  Others processes would have different spectral properties. For instance, the particle energy distribution related to magnetic reconnection would have an intrinsic spectral break normally ocurring between optical and radio bands. 
However, the current observational sample fails to distinguish between different models, as at different wavelengths (e.g., radio and X-rays), spatially consistent PWN structures  are rarely detected.

\begin{figure}[ht]
    \centering
    \includegraphics[width=1\linewidth]{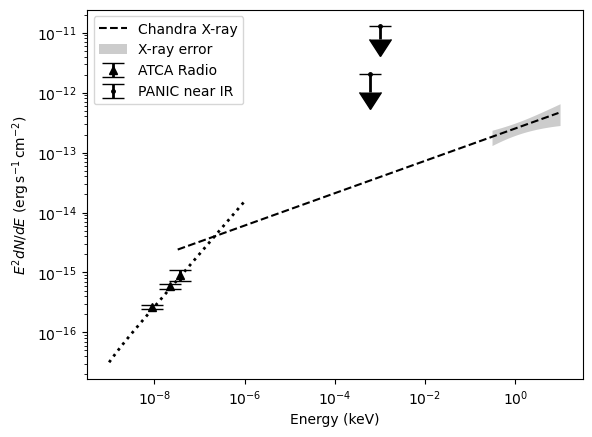}
    \caption{Multi-band SED in the radio jet region, the triangle error bars and gray belt indicate to the radio and X-ray spectra in our studies; and the dashed and dotted lines extrapolate the X-ray and radio synchrotron spectra; the upper limits show \textit{J} and \textit{K} band fluxes in near-Infrared observations.   }
    \label{fig:SED}
\end{figure}

Thanks to the unique radio and X-ray spatial consistency in the SW jet of \gname\ PWN, we can directly compare the observed flux levels to create a broad band SYN spectrum energy distribution (SED), and this is shown in Figure \ref{fig:SED}. 
Our observed radio spectrum (at 3, 6, and 16\,cm) as well as the 4-7.5\,keV \emph{Chandra} X-ray non-thermal spectrum of the SW jet are shown.  
In addition, we searched for observations in other bands to better constrain the SED. Unfortunately, few archival observations were found to detect the PWN successfully. The \emph{WISE} 22\,$\mu$m observation and P band VLA observation show the structure of the SNR, but unfortunately failed to resolve the central PWN, partially due to limited resolution. Also at P band,  the steeper shell spectrum obscures the flatter PWN at low frequencies \citep{2010AJ....140.1868W,2003ApJ...588..992R}. Here we also use upper limits from the  \textit{J} and \textit{K} band optical observations from PANIC (Persson’s Auxiliary Nasmyth Infrared Camera) close to the pulsar that were scaled to the region we measure in radio and X-rays \citep{2006ApJ...644.1056K}.

As is mentioned above, the jet has a radio spectral index $\alpha\sim-0.10$ and an X-ray photon index $\Gamma\sim1.73$, suggesting a spectral break $\Delta\alpha\sim 0.6$. The extrapolated radio and X-ray spectra intersect at a frequency between 50 and 100\,GHz, which generally aligns with expectations. The PANIC observation unfortunately fails to strongly constraining the SED in the near IR bands. 
In any case, the SYN SED of the SW jet region is not a single power law, and must either slowly or sharply break somewhere. Future observations at millimeter wavelengths would be useful to explore the transition region and accurately locate the breaking point. 
Here we also note that the radio spectrum of the jet is flatter than the previous VLA spectral index measurement of the entire radio PWN region ($\alpha_p \simeq -0.25$), which may have been hinted in the spectral tomography maps \citep{2002ApJ...572..202T}. For our observation, the missing flux problem at short $u-v$ spacings shold not be a significant issue for the small-scaled torus and jet (20\arcsec) studied here, as it is for the SNR shell and may even be for the PWN as a whole. The jet region could easily have a flatter radio spectrum at the small scales considered here, as the previous result focused on a much larger region and did not differentiate the radio torus and jet from the larger features more distant from the pulsar. 

Here we consider a simple broken power law SYN spectrum in the PWN jet.
Observationally, such a spectral feature could be formed due to the synchrotron cooling effect, because of  the unequal energy loss of electrons at different energies. However, we do not expect significant cooling in the case of the jet region below the X-ray band.
The cooling effect should only become significant further away from the pulsar, and comparisons of radio and X-ray PWNe often show anti-correlated features at large scales \citep[e.g., B1706, Vela X, and Snail;][]{2023ApJ...945...82L,2003MNRAS.343..116D,2016ApJ...820..100M}. This anti-correlation is suggested to be driven by both particle accumulation and synchrotron cooling, where dilute particles spreads out and accumulates in the outer regions where they cool down to emit only radio emissions \citep{2008ApJ...684..542K}. However in our case, we expect to predominately have particles that are freshly injected close to the pulsar jet.
It is expected that particles injected into a  PWN  follow a relationship between age ($\tau$), $B$-field strength ($B$), and breaking frequency ($\nu_b$):
\[\tau = [\frac{\nu_b}{10^{21}\,\text{Hz}}]^{-1/2}[\frac{B}{1\,\mu\text{G}}]^{-3/2}\,\text{kyr}.\]
Assuming a cooling time $\tau$ of 2\,kyr and a break around $10^{11}$\,Hz (see Figure \ref{fig:SED}), it would require a $B$-field stronger than 1\,mG, much higher than our measurement \citep{2006ARA&A..44...17G}.
We therefore suggest that the spectral break is not due to the cooling effect, but is likely an intrinsic spectral feature. 

An intrinsic broken power law synchrotron spectrum can be produced from magnetic reconnection of a striped pulsar wind where magnetic energies are transferred to particles gradually to form a flat power law energy distribution with a index $p_1 \simeq -1.5$; then the highest energy particles could pass the shock and gain energy from Fermi acceleration, forming another power-law component with index $p_2\simeq -2.5$ at higher energies \citep{2011ApJ...726...75S}.  
For our case of \gname\ jet, we obtain a $p_1 \simeq -1.2$ below the break and $p_2 \simeq -2.5$ at high energies, as well as a break between radio and optical bands; all these generally align with the expectation of the model of magnetic reconnection.

We note that the current multi-band spectrum may also imply other energy distributions like Maxwellian plus a power law, which could be interpreted with theories like magnetic reconnection or Weibel instability \citep{2011ApJ...726...75S,1959PhysRevLett.2.83W}. This could  be investigated with future multi-band observations.

\subsection{PWN Geometry and Evolution}

\begin{figure}
    \centering
    \includegraphics[width=1.0\linewidth]{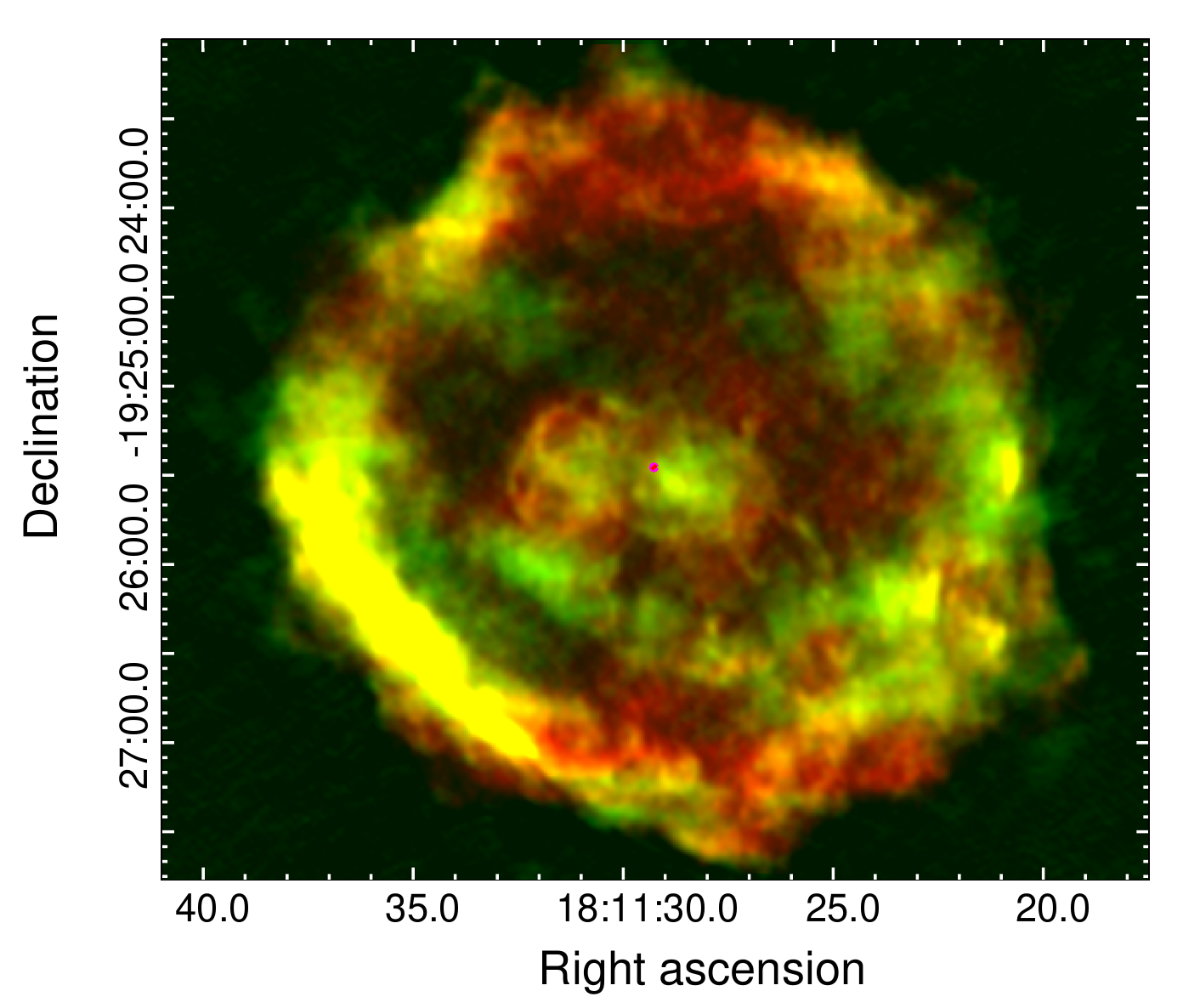}\\
    \caption{Comparison of the previous 3\,cm VLA observation and the new ATCA 3\,cm observation in \gname\ SNR region. The top panel shows a RGB map comparing the old (red) and new (green) images of the whole SNR, the magenta circle shows the position of the pulsar.   }
    \label{fig:snr_expand}
\end{figure}

\begin{figure}
    \centering
    \includegraphics[width=1.0\linewidth]{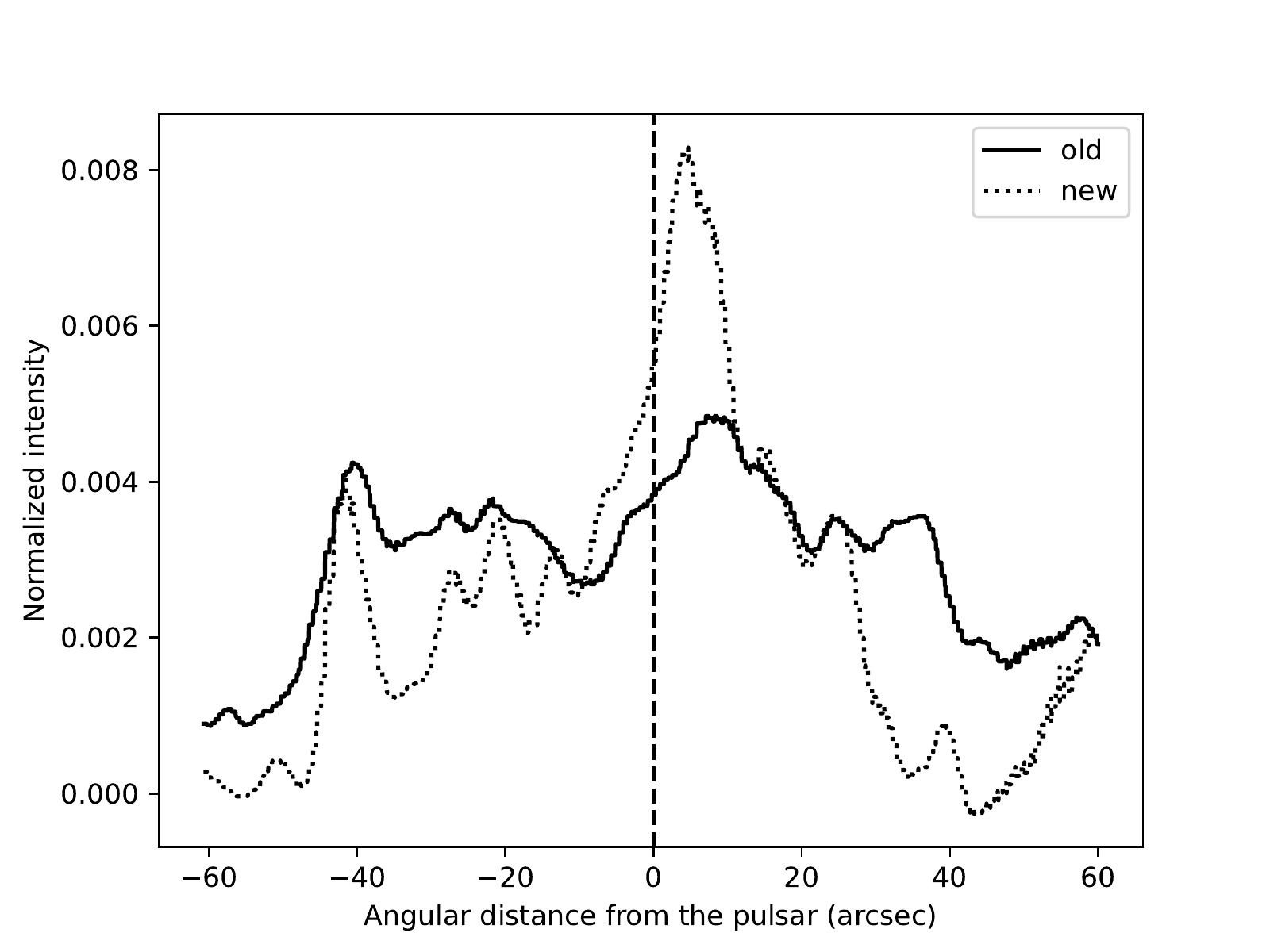}
    \caption{The normalized intensity profile along the jet direction \citep[see][]{2016ApJ...819..160B} in the old and new observations.}
    \label{fig:jet-dyna}
\end{figure}

\begin{figure}
    \centering    
    \includegraphics[width=1.0\linewidth]{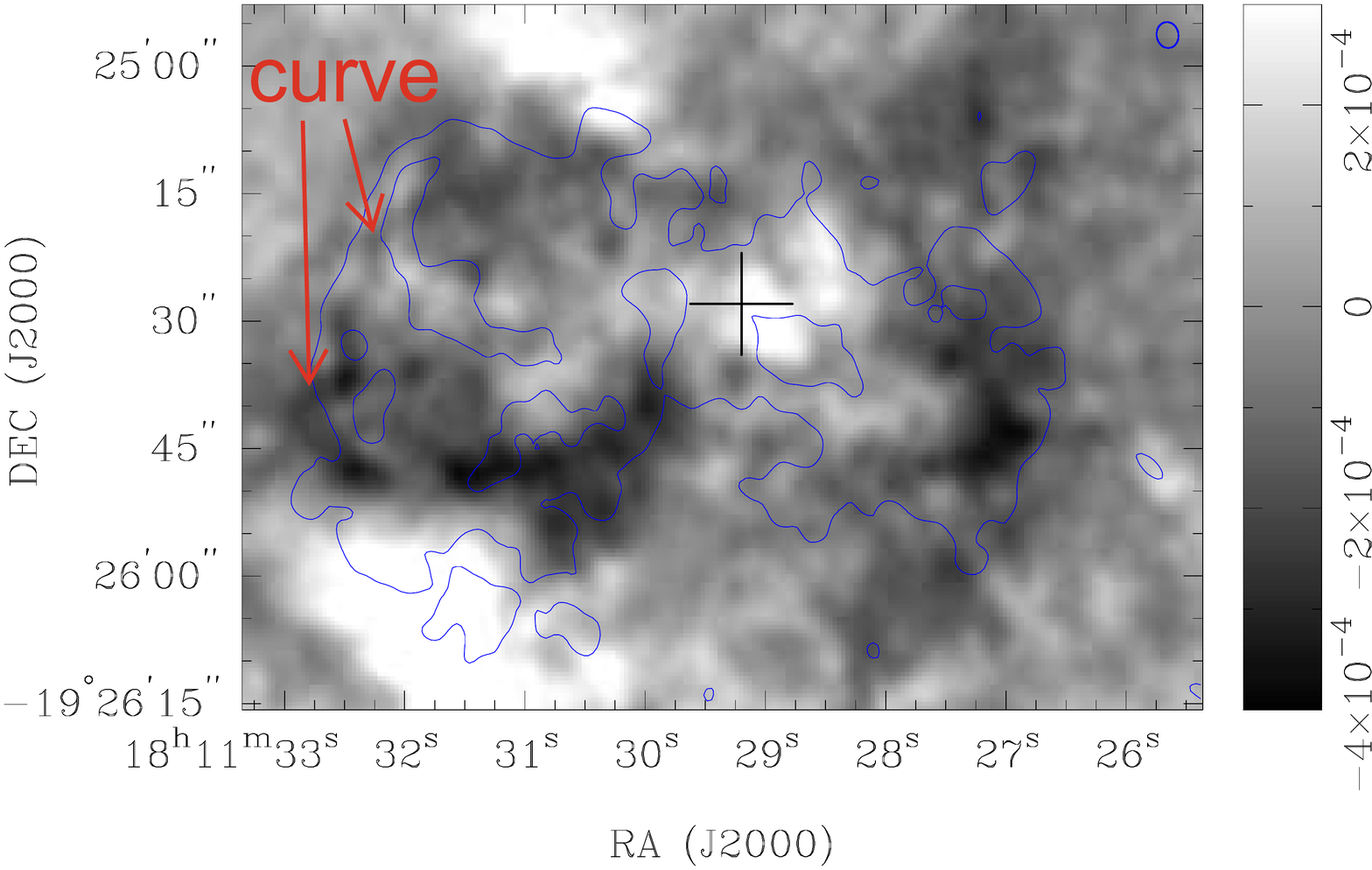}
    \caption{The residual of the new observation subtracting the old one to suggest differences of PWN structures over 20 years, and the overlapping blue contours shows the PWN at 0.1 and 0.144\mJyPerBeam in the previous VLA observation. The red arrows mark the curve showing displacements of PWN structures. }
    \label{fig:PWN-dyna}
\end{figure}

Previous VLA observations in 2001 at 3\,cm provided a mosaic image with a FWHM beam size of $3.07\arcsec\times2.56\arcsec$ and position angle of $9.608^\circ$, which can clearly resolve detailed SNR and PWN structures like the torus and jet \citep{2003ApJ...588..992R}. 
We produce a new ATCA 3\,cm image with the same FWHM and compare with the VLA image to show motions of structures inside the SNR as shown in Figure \ref{fig:snr_expand}.
The new observations (in green) show morphological coincidence with the old ones (in red). The SNR rim and PWN do not exhibit large geometrical differences between the two observations. The old observation shows rather uniform background emission within the SNR shell, while the new observation shows more clumpy structures inside region (e.g., the linear filament in the SE part hinting at a reverse shock). 
We suggest this could be due to different short baseline coverage. Supplementary observations might be helpful to confirm these structures in the new observations.
Within the PWN region, small structural details (e.g., torus/jet, arcs) are resolved in both and align with each other.
The \texttt{imdiff} task in MIRIAD package suggests that the SNR is expanding over 20 years, and returns a expansion rate of $0.4\%$ for the SNR.
This implies a expansion of $0\arcsec.05$\,year$^{-1}$ given the SNR scale of $\sim250\arcsec$, this estimate generally comparable the previous result comparing data between 1984 and 2001 \citep{2003ApJ...598L..27T}.

Due to the rather complicated SNR geometries and limited resolution, we are unable to accurately measure every motion of nebular structures close to the pulsar (e.g., jet).
However, we show in Figure \ref{fig:jet-dyna} the intensity profile of the new and old radio observations along the direction of the jet as mentioned in a previous study \citep{2003ApJ...588..992R}, and this reveals  the
volatility of the bright spots is a feature of both the X-ray and radio jet. %expanding trend of the PWN.  Along the jet, it isn't clear there is expansion. 

In Figure \ref{fig:PWN-dyna}, the residual map obtained by \texttt{imdiff} also suggests the motion of some other PWN structures. 
We subtract the old 3\,cm observation of the PWN from the new one, with an amplitude scale fixed at 0.5 to better resolve the displacements. 
A curve, as is marked in Fig \ref{fig:PWN-dyna}, is detected along the rim of the eastern arc, suggesting potential motions of this structure. 
An average width of this curve is around 2.4\arcsec (comparable to a beam size), this implies that the arc region is floating with a putative velocity of \textless2700\,km/s. Interestingly, since relativistic motion has been observed in the X-ray SW jet \citep{2003ApJ...588..992R}, then it is rational to detect high speeds even at the end of its PWN jets.
Both Figure \ref{fig:jet-dyna} and \ref{fig:PWN-dyna} hints at an overall expanding tendency of the PWN structures, which might indicate the reverse shock has either already passed or has not yet reached the PWN. 
It is also notable that the bright spot in the jet shows possible change in both position and brightness in these Figures.
\emph{Chandra} X-ray observations discovered some putative morphological changes of the innermost spot from 2000 to 2013, these might show connections with the differences discovered in the radio jet.

\begin{figure}
    \centering
    \includegraphics[width=1.0\linewidth]{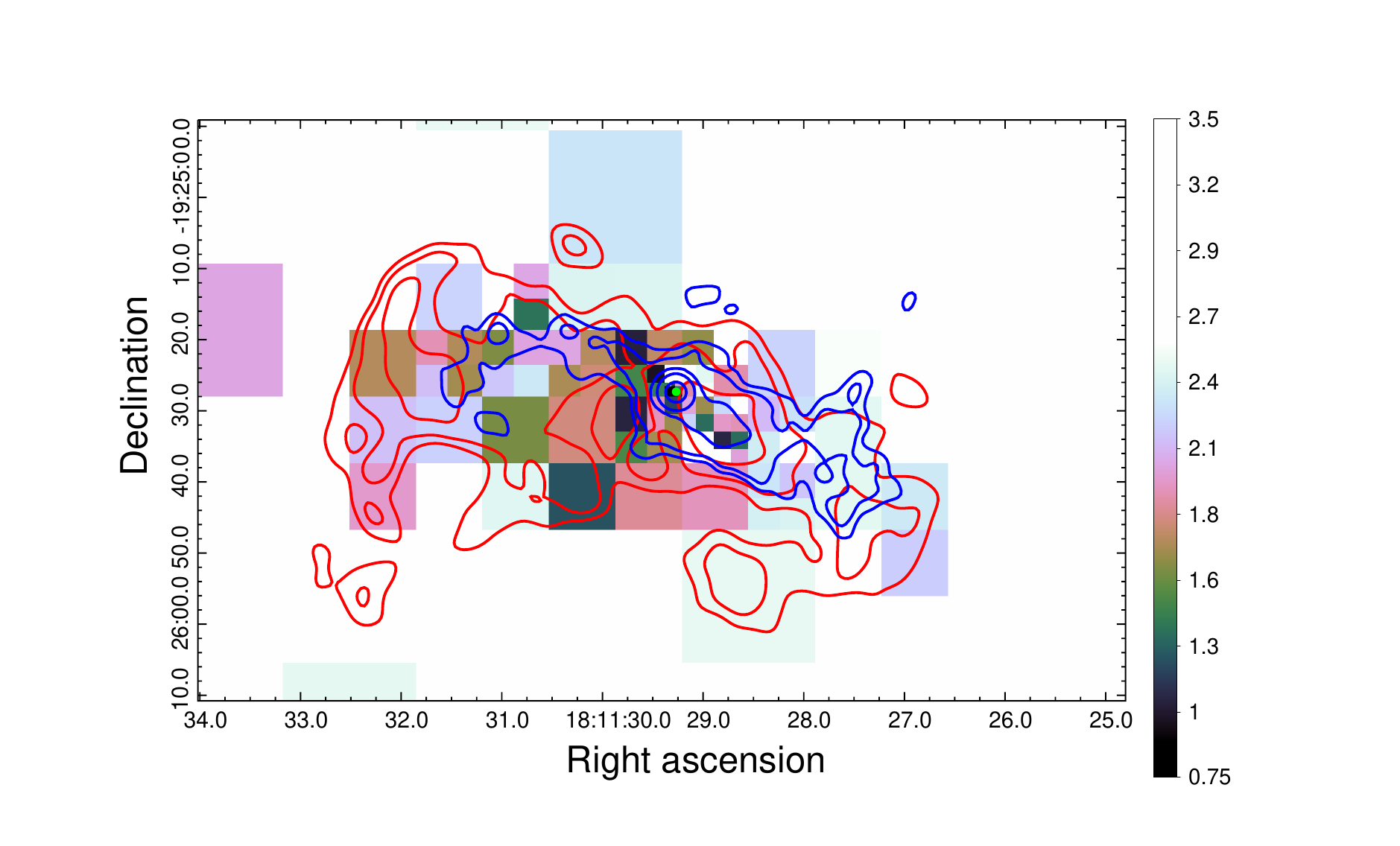}
    \caption{Non-thermal photon index map of \gname\ PWN, overlapping with \emph{Chandra} X-ray map (blue contours) and 6\,cm ATCA map (red contour); the color bar on the right shows the local photon index. }
    \label{fig:X-ray_index}
\end{figure}

To better analyze the nature of the PWN region, we obtained the 4-7.5\,keV \emph{Chandra} X-ray photon index ($\Gamma$) map. 
We use the quad-tree adaptive binning task \texttt{dmnautilus} in CIAO and then extract photon indices of non-thermal spectra at local bins; all these are shown in Figure \ref{fig:X-ray_index}. 
On large scales, a hard spectrum is generally detected across the PWN region, with a general softening towards the oute regions.
This map also show some detailed features aligning with the X-ray PWN distribution.
Hard X-ray spectrum ($\Gamma$ < 2) is discovered in both SW and NE jet region close to the pulsar, while it is not as hard in the equatorial region. 
Interestingly, there is photon index ($\sim1.8$) where the NE jet is turn to SE, this suggests that the belted jet is of similar component with the NE jet.
However in the position of radio arc (i.e., X-ray jet inflecting point), the index goes softer than 2.1, this hints a scenario of possible compression between the jet and SN ejecta. 
It is remarkable that an X-ray thermal clump is found beyond the NE jet in Figure \ref{fig:radioX}, and the radio arc locates just between the X-ray jet and clump. 
We here suggest that the relativistic plasma in jet might interact with the SNR ejecta at this region and turns to SE.
Some other radio observations (e.g., Crab, Snail) also show filaments nearby the interface of these two different materials (due to Rayleigh-Taylor instabilities), which are also discovered in the arc region; meanwhile, the outer NW jet region seems to be expanding rapidly (see above). All these could support such a possible interaction between the jet and SNR ejecta.  
\section{Conclusions}
\label{sec:conclusion}

We present a study of \gname\ PWN using new ATCA observations at 3 and 6\,cm in 2022, as well as at 16\,cm  in 2024. 
The main points are summarized below: 
\begin{enumerate}
    \item Our high resolution observations clearly resolved detailed structures inside the SNR and PWN region. The radio PWN interestingly show a torus/jet feature as in X-ray observations. Complicated structures (e.g., the arc) are shown farther away from the pulsar, and we compared the radio and X-ray emissions to show some possible spatial connections.
    \item The multi-band spectrum of the jet is less-likely to be a simple power-law, this possibly indicates complex particle acceleration mechanisms such as magnetic reconnection. 
    \item A Helical $B$-field is found in the PWN jet region, and the local magnetic field strength is measured to be around 85\,$\mu$G.\\ 
\end{enumerate}
To better understand the spectral properties of this PWN, better frequency coverage is essential for accurately constraining the SED and following numerical simulations. Follow-up observations at higher frequencies in the future may help determine the break energy and test scenarios like magnetic reconnection.  In addition, the $B$-field configuration of the detailed structures provides important information for MHD simulations of PWN. \\

C.-Y.N. is supported by a GRF grant of the Hong Kong Government under HKU 17304524. This work is also supported by the National Natural Science Foundation of China (NSFC) grants 12261141691 and the Fundamental Research Funds for the Central Universities, Sun Yat-sen University, No. 24qnpy123. The Australia Telescope Compact Array is part of the Australia Telescope National Facility which is funded by the Commonwealth of Australia for operation as a National Facility managed by CSIRO. This research has made use of data obtained from the Chandra Data Archive provided by the \emph{Chandra} X-ray Center (CXC). We also thank Zijian Qiu and Long Ji for helps in X-ray data reduction.  This study also makes use of software in the application packages Miriad, CIAO, and Sherpa.

\appendix 
\section{Depolarization function}
\label{sec:depol}
For a radio PL observation of a source at rotation measure of $RM$ with a half bandwidth of $f_b$ and frequencies and centered at $f_c$. According to Equation \ref{eq:faraday}, if we assume that polarization angle rotating across frequencies is quasi-symmetric about the PL direction at f$_c$, a half angle $\frac{\Delta \chi}{2}$ of the PL direction coverage should be
\begin{equation}
    \frac{\Delta \chi}{2} = \chi_{c-b} - \chi_{c} = RM\cdot(\frac{c^2}{(f_c-f_b)^2} - \frac{c^2}{f_c^2}) \\
    % =  RM\cdot c^2 \cdot \frac{f_c^2 - (f_c-f_b)^2}{(f_c-f_b)^2 f_c^2} 
    = RM\cdot c^2 \cdot \frac{2\frac{f_c}{f_b} - 1}{(\frac{f_c}{f_b}-1)^2 f_c^2}. 
\end{equation}

Next, we note that a PL emission at frequency $f$ should have a symmetric PL emission at frequency $2f_c-f$ about the $\chi_c$; therefore, a remained PL fraction after the bandwidth depolarization is: 
\begin{equation}
    PF = \left|\int^{\frac{\Delta \chi}{2}}_0 \cos{2\theta} \mbox{d}\theta / \frac{\Delta \chi}{2}\right| =  \left|\sin(\Delta \chi)/ \Delta \chi\right|
\end{equation}
and then: 
\begin{equation}
    DPF = 1- PF = 1 - \left|\sin(\Delta \chi)/ \Delta \chi\right|. 
\end{equation}
Under the case of our 16\,cm observation, we obtained a $\left|\frac{\Delta \chi}{2}\right|$ of $\sim 41.5$\,rad, and then suggest a depolarization fraction $DPF$ of 0.99, so that the 16\,cm observation has been heavily depolarized.

\bibliography{sample631}{}

\begin{thebibliography}{}
\expandafter\ifx\csname natexlab\endcsname\relax\def\natexlab#1{#1}\fi
\providecommand{\url}[1]{\href{#1}{#1}}
\providecommand{\dodoi}[1]{doi:~\href{http://doi.org/#1}{\nolinkurl{#1}}}
\providecommand{\doeprint}[1]{\href{http://ascl.net/#1}{\nolinkurl{http://ascl.net/#1}}}
\providecommand{\doarXiv}[1]{\href{https://arxiv.org/abs/#1}{\nolinkurl{https://arxiv.org/abs/#1}}}

\bibitem[{{Asada} {et~al.}(2002){Asada}, {Inoue}, {Uchida}, {Kameno}, {Fujisawa}, {Iguchi}, \& {Mutoh}}]{2002PASJ...54L..39A}
{Asada}, K., {Inoue}, M., {Uchida}, Y., {et~al.} 2002, \pasj, 54, L39, \dodoi{10.1093/pasj/54.3.L39}

\bibitem[{{Atoyan}(1999)}]{1999A&A...346L..49A}
{Atoyan}, A.~M. 1999, \aap, 346, L49, \dodoi{10.48550/arXiv.astro-ph/9905204}

\bibitem[{{Becker} {et~al.}(1985){Becker}, {Markert}, \& {Donahue}}]{1985ApJ...296..461B}
{Becker}, R.~H., {Markert}, T., \& {Donahue}, M. 1985, \apj, 296, 461, \dodoi{10.1086/163465}

\bibitem[{{Bietenholz}(2006)}]{2006ApJ...645.1180B}
{Bietenholz}, M.~F. 2006, \apj, 645, 1180, \dodoi{10.1086/504584}

\bibitem[{{Bietenholz} {et~al.}(2001){Bietenholz}, {Frail}, \& {Hester}}]{2001ApJ...560..254B}
{Bietenholz}, M.~F., {Frail}, D.~A., \& {Hester}, J.~J. 2001, \apj, 560, 254, \dodoi{10.1086/322244}

\bibitem[{{Bogovalov} \& {Khangoulyan}(2002)}]{2002AstL...28..373B}
{Bogovalov}, S.~V., \& {Khangoulyan}, D.~V. 2002, Astronomy Letters, 28, 373, \dodoi{10.1134/1.1484137}

\bibitem[{{Borkowski} {et~al.}(2016){Borkowski}, {Reynolds}, \& {Roberts}}]{2016ApJ...819..160B}
{Borkowski}, K.~J., {Reynolds}, S.~P., \& {Roberts}, M. S.~E. 2016, \apj, 819, 160, \dodoi{10.3847/0004-637X/819/2/160}

\bibitem[{{Briggs}(1995)}]{1995AAS...18711202B}
{Briggs}, D.~S. 1995, in American Astronomical Society Meeting Abstracts, Vol. 187, American Astronomical Society Meeting Abstracts, 112.02

\bibitem[{{Bucciantini} {et~al.}(2023){Bucciantini}, {Ferrazzoli}, {Bachetti}, {Rankin}, {Di Lalla}, {Sgr{\`o}}, {Omodei}, {Kitaguchi}, {Mizuno}, {Gunji}, {Watanabe}, {Baldini}, {Slane}, {Weisskopf}, {Romani}, {Possenti}, {Marshall}, {Silvestri}, {Pacciani}, {Negro}, {Muleri}, {de O{\~n}a Wilhelmi}, {Xie}, {Heyl}, {Pesce-Rollins}, {Wong}, {Pilia}, {Agudo}, {Antonelli}, {Baumgartner}, {Bellazzini}, {Bianchi}, {Bongiorno}, {Bonino}, {Brez}, {Capitanio}, {Castellano}, {Cavazzuti}, {Chen}, {Ciprini}, {Costa}, {De Rosa}, {Del Monte}, {Di Gesu}, {Di Marco}, {Donnarumma}, {Doroshenko}, {Dov{\v{c}}iak}, {Ehlert}, {Enoto}, {Evangelista}, {Fabiani}, {Garcia}, {Hayashida}, {Iwakiri}, {Jorstad}, {Kaaret}, {Karas}, {Kislat}, {Kolodziejczak}, {Krawczynski}, {La Monaca}, {Latronico}, {Liodakis}, {Maldera}, {Manfreda}, {Marin}, {Marinucci}, {Marscher}, {Massaro}, {Matt}, {Mitsuishi}, {Ng}, {O'Dell}, {Oppedisano}, {Papitto}, {Pavlov}, {Peirson}, {Perri}, {Petrucci}, {Poutanen}, {Puccetti}, {Ramsey}, {Ratheesh}, {Roberts},
  {Soffitta}, {Spandre}, {Swartz}, {Tamagawa}, {Tavecchio}, {Taverna}, {Tawara}, {Tennant}, {Thomas}, {Tombesi}, {Trois}, {Tsygankov}, {Turolla}, {Vink}, {Wu}, \& {Zane}}]{2023NatAs...7..602B}
{Bucciantini}, N., {Ferrazzoli}, R., {Bachetti}, M., {et~al.} 2023, Nature Astronomy, 7, 602, \dodoi{10.1038/s41550-023-01936-8}

\bibitem[{{Cao} {et~al.}(2021){Cao}, {Aharonian}, {An}, {Axikegu}, {Bai}, {Bao}, {Bastieri}, {Bi}, {Bi}, {Cai}, {Cai}, {Cao}, {Chang}, {Chang}, {Chang}, {Chen}, {Chen}, {Chen}, {Chen}, {Chen}, {Chen}, {Chen}, {Chen}, {Chen}, {Chen}, {Chen}, {Chen}, {Chen}, {Cheng}, {Cheng}, {Cui}, {Cui}, {Cui}, {Dai}, {Dai}, {Dai}, {Danzengluobu}, {della Volpe}, {D'Ettorre Piazzoli}, {Dong}, {Fan}, {Fan}, {Fan}, {Fang}, {Fang}, {Feng}, {Feng}, {Feng}, {Feng}, {Gao}, {Gao}, {Gao}, {Gao}, {Ge}, {Geng}, {Gong}, {Gou}, {Gu}, {Guo}, {Guo}, {Guo}, {Guo}, {Han}, {He}, {He}, {He}, {He}, {He}, {He}, {Heller}, {Hor}, {Hou}, {Hou}, {Hu}, {Hu}, {Hu}, {Hu}, {Huang}, {Huang}, {Huang}, {Huang}, {Huang}, {Ji}, {Ji}, {Jia}, {Jiang}, {Jiang}, {Jin}, {Kuleshov}, {Levochkin}, {Li}, {Li}, {Li}, {Li}, {Li}, {Li}, {Li}, {Li}, {Li}, {Li}, {Li}, {Li}, {Li}, {Li}, {Li}, {Li}, {Li}, {Liang}, {Liang}, {Lin}, {Liu}, {Liu}, {Liu}, {Liu}, {Liu}, {Liu}, {Liu}, {Liu}, {Liu}, {Liu}, {Liu}, {Liu}, {Liu}, {Liu}, {Liu}, {Long}, {Lu}, {Lv}, {Ma}, {Ma}, {Ma},
  {Mao}, {Masood}, {Mitthumsiri}, {Montaruli}, {Nan}, {Pang}, {Pattarakijwanich}, {Pei}, {Qi}, {Ruffolo}, {Rulev}, {S{\'a}iz}, {Shao}, {Shchegolev}, {Sheng}, {Shi}, {Song}, {Stenkin}, {Stepanov}, {Sun}, {Sun}, {Sun}, {Tam}, {Tang}, {Tian}, {Wang}, {Wang}, {Wang}, {Wang}, {Wang}, {Wang}, {Wang}, {Wang}, {Wang}, {Wang}, {Wang}, {Wang}, {Wang}, {Wang}, {Wang}, {Wang}, {Wang}, {Wang}, {Wang}, {Wang}, {Wang}, {Wei}, {Wei}, {Wei}, {Wen}, {Wu}, {Wu}, {Wu}, {Wu}, {Wu}, {Xi}, {Xia}, {Xia}, {Xiang}, {Xiao}, {Xiao}, {Xin}, {Xin}, {Xing}, {Xu}, {Xu}, {Xue}, {Yan}, {Yang}, {Yang}, {Yang}, {Yang}, {Yang}, {Yang}, {Yang}, {Yao}, {Yao}, {Ye}, {Yin}, {Yin}, {You}, {You}, {Yu}, {Yuan}, {Zeng}, {Zeng}, {Zeng}, {Zeng}, {Zha}, {Zhai}, {Zhang}, {Zhang}, {Zhang}, {Zhang}, {Zhang}, {Zhang}, {Zhang}, {Zhang}, {Zhang}, {Zhang}, {Zhang}, {Zhang}, {Zhang}, {Zhang}, {Zhang}, {Zhang}, {Zhang}, {Zhang}, {Zhang}, {Zhao}, {Zhao}, {Zhao}, {Zhao}, {Zhao}, {Zheng}, {Zheng}, {Zhou}, {Zhou}, {Zhou}, {Zhou}, {Zhou}, {Zhou}, {Zhu}, {Zhu}, {Zhu},
  {Zhu}, \& {Zuo}}]{2021Natur.594...33C}
{Cao}, Z., {Aharonian}, F.~A., {An}, Q., {et~al.} 2021, \nat, 594, 33, \dodoi{10.1038/s41586-021-03498-z}

\bibitem[{{Dodson} {et~al.}(2003){Dodson}, {Lewis}, {McConnell}, \& {Deshpande}}]{2003MNRAS.343..116D}
{Dodson}, R., {Lewis}, D., {McConnell}, D., \& {Deshpande}, A.~A. 2003, \mnras, 343, 116, \dodoi{10.1046/j.1365-8711.2003.06653.x}

\bibitem[{{Dubner} {et~al.}(2017){Dubner}, {Castelletti}, {Kargaltsev}, {Pavlov}, {Bietenholz}, \& {Talavera}}]{2017ApJ...840...82D}
{Dubner}, G., {Castelletti}, G., {Kargaltsev}, O., {et~al.} 2017, \apj, 840, 82, \dodoi{10.3847/1538-4357/aa6983}

\bibitem[{{Fuentes} {et~al.}(2023){Fuentes}, {G{\'o}mez}, {Mart{\'\i}}, {Perucho}, {Zhao}, {Lico}, {Lobanov}, {Bruni}, {Kovalev}, {Chael}, {Akiyama}, {Bouman}, {Sun}, {Cho}, {Traianou}, {Toscano}, {Dahale}, {Foschi}, {Gurvits}, {Jorstad}, {Kim}, {Marscher}, {Mizuno}, {Ros}, \& {Savolainen}}]{2023NatAs...7.1359F}
{Fuentes}, A., {G{\'o}mez}, J.~L., {Mart{\'\i}}, J.~M., {et~al.} 2023, Nature Astronomy, 7, 1359, \dodoi{10.1038/s41550-023-02105-7}

\bibitem[{{Gaensler} \& {Slane}(2006)}]{2006ARA&A..44...17G}
{Gaensler}, B.~M., \& {Slane}, P.~O. 2006, \araa, 44, 17, \dodoi{10.1146/annurev.astro.44.051905.092528}

\bibitem[{Giacinti \& Kirk(2019)}]{giacinti2019electronaccelerationcrabnebula}
Giacinti, G., \& Kirk, J.~G. 2019, Electron Acceleration in the Crab Nebula.
\newblock \doarXiv{1909.06230}

\bibitem[{{Govoni} \& {Feretti}(2004)}]{2004IJMPD..13.1549G}
{Govoni}, F., \& {Feretti}, L. 2004, International Journal of Modern Physics D, 13, 1549, \dodoi{10.1142/S0218271804005080}

\bibitem[{{Kaplan} \& {Moon}(2006)}]{2006ApJ...644.1056K}
{Kaplan}, D.~L., \& {Moon}, D.-S. 2006, \apj, 644, 1056, \dodoi{10.1086/503794}

\bibitem[{{Kargaltsev} {et~al.}(2008){Kargaltsev}, {Misanovic}, {Pavlov}, {Wong}, \& {Garmire}}]{2008ApJ...684..542K}
{Kargaltsev}, O., {Misanovic}, Z., {Pavlov}, G.~G., {Wong}, J.~A., \& {Garmire}, G.~P. 2008, \apj, 684, 542, \dodoi{10.1086/589145}

\bibitem[{{Kargaltsev} \& {Pavlov}(2008)}]{2008AIPC..983..171K}
{Kargaltsev}, O., \& {Pavlov}, G.~G. 2008, in American Institute of Physics Conference Series, Vol. 983, 40 Years of Pulsars: Millisecond Pulsars, Magnetars and More, ed. C.~{Bassa}, Z.~{Wang}, A.~{Cumming}, \& V.~M. {Kaspi} (AIP), 171--185, \dodoi{10.1063/1.2900138}

\bibitem[{{Komissarov} \& {Lyubarsky}(2003)}]{2003MNRAS.344L..93K}
{Komissarov}, S.~S., \& {Lyubarsky}, Y.~E. 2003, \mnras, 344, L93, \dodoi{10.1046/j.1365-8711.2003.07097.x}

\bibitem[{{Kothes} {et~al.}(2008){Kothes}, {Landecker}, {Reich}, {Safi-Harb}, \& {Arzoumanian}}]{2008ApJ...687..516K}
{Kothes}, R., {Landecker}, T.~L., {Reich}, W., {Safi-Harb}, S., \& {Arzoumanian}, Z. 2008, \apj, 687, 516, \dodoi{10.1086/591653}

\bibitem[{{Kothes} \& {Reich}(2001)}]{2001A&A...372..627K}
{Kothes}, R., \& {Reich}, W. 2001, \aap, 372, 627, \dodoi{10.1051/0004-6361:20010407}

\bibitem[{{Kothes} {et~al.}(2006){Kothes}, {Reich}, \& {Uyan{\i}ker}}]{2006ApJ...638..225K}
{Kothes}, R., {Reich}, W., \& {Uyan{\i}ker}, B. 2006, \apj, 638, 225, \dodoi{10.1086/498666}

\bibitem[{{Liu} {et~al.}(2023){Liu}, {Ng}, \& {Dodson}}]{2023ApJ...945...82L}
{Liu}, Y.~H., {Ng}, C.~Y., \& {Dodson}, R. 2023, \apj, 945, 82, \dodoi{10.3847/1538-4357/acb20d}

\bibitem[{{Ma} {et~al.}(2016){Ma}, {Ng}, {Bucciantini}, {Slane}, {Gaensler}, \& {Temim}}]{2016ApJ...820..100M}
{Ma}, Y.~K., {Ng}, C.~Y., {Bucciantini}, N., {et~al.} 2016, \apj, 820, 100, \dodoi{10.3847/0004-637X/820/2/100}

\bibitem[{{Ng} \& {Romani}(2004)}]{2004ApJ...601..479N}
{Ng}, C.~Y., \& {Romani}, R.~W. 2004, \apj, 601, 479, \dodoi{10.1086/380486}

\bibitem[{{Ng} \& {Romani}(2008)}]{2008ApJ...673..411N}
---. 2008, \apj, 673, 411, \dodoi{10.1086/523935}

\bibitem[{{Pacholczyk}(1970)}]{1970ranp.book.....P}
{Pacholczyk}, A.~G. 1970, {Radio astrophysics. Nonthermal processes in galactic and extragalactic sources}

\bibitem[{{Porth} {et~al.}(2014){Porth}, {Komissarov}, \& {Keppens}}]{2014MNRAS.438..278P}
{Porth}, O., {Komissarov}, S.~S., \& {Keppens}, R. 2014, \mnras, 438, 278, \dodoi{10.1093/mnras/stt2176}

\bibitem[{{Roberts} {et~al.}(2003){Roberts}, {Tam}, {Kaspi}, {Lyutikov}, {Vasisht}, {Pivovaroff}, {Gotthelf}, \& {Kawai}}]{2003ApJ...588..992R}
{Roberts}, M. S.~E., {Tam}, C.~R., {Kaspi}, V.~M., {et~al.} 2003, \apj, 588, 992, \dodoi{10.1086/374266}

\bibitem[{{Sault} {et~al.}(1995){Sault}, {Teuben}, \& {Wright}}]{1995ASPC...77..433S}
{Sault}, R.~J., {Teuben}, P.~J., \& {Wright}, M.~C.~H. 1995, in Astronomical Society of the Pacific Conference Series, Vol.~77, Astronomical Data Analysis Software and Systems IV, ed. R.~A. {Shaw}, H.~E. {Payne}, \& J.~J.~E. {Hayes}, 433, \dodoi{10.48550/arXiv.astro-ph/0612759}

\bibitem[{{Sironi} \& {Spitkovsky}(2011)}]{2011ApJ...726...75S}
{Sironi}, L., \& {Spitkovsky}, A. 2011, \apj, 726, 75, \dodoi{10.1088/0004-637X/726/2/75}

\bibitem[{{Slane}(2017)}]{2017hsn..book.2159S}
{Slane}, P. 2017, in Handbook of Supernovae, ed. A.~W. {Alsabti} \& P.~{Murdin}, 2159, \dodoi{10.1007/978-3-319-21846-5_95}

\bibitem[{{Slane} {et~al.}(2004){Slane}, {Helfand}, {van der Swaluw}, \& {Murray}}]{2004ApJ...616..403S}
{Slane}, P., {Helfand}, D.~J., {van der Swaluw}, E., \& {Murray}, S.~S. 2004, \apj, 616, 403, \dodoi{10.1086/424814}

\bibitem[{{Tam} \& {Roberts}(2003)}]{2003ApJ...598L..27T}
{Tam}, C., \& {Roberts}, M. S.~E. 2003, \apjl, 598, L27, \dodoi{10.1086/380557}

\bibitem[{{Tam} {et~al.}(2002){Tam}, {Roberts}, \& {Kaspi}}]{2002ApJ...572..202T}
{Tam}, C., {Roberts}, M. S.~E., \& {Kaspi}, V.~M. 2002, \apj, 572, 202, \dodoi{10.1086/340229}

\bibitem[{{Torii} {et~al.}(1999){Torii}, {Tsunemi}, {Dotani}, {Mitsuda}, {Kawai}, {Kinugasa}, {Saito}, \& {Shibata}}]{1999ApJ...523L..69T}
{Torii}, K., {Tsunemi}, H., {Dotani}, T., {et~al.} 1999, \apjl, 523, L69, \dodoi{10.1086/312251}

\bibitem[{Weibel(1959)}]{1959PhysRevLett.2.83W}
Weibel, E.~S. 1959, Phys. Rev. Lett., 2, 83, \dodoi{10.1103/PhysRevLett.2.83}

\bibitem[{{Wilson} {et~al.}(2011){Wilson}, {Ferris}, {Axtens}, {Brown}, {Davis}, {Hampson}, {Leach}, {Roberts}, {Saunders}, {Koribalski}, {Caswell}, {Lenc}, {Stevens}, {Voronkov}, {Wieringa}, {Brooks}, {Edwards}, {Ekers}, {Emonts}, {Hindson}, {Johnston}, {Maddison}, {Mahony}, {Malu}, {Massardi}, {Mao}, {McConnell}, {Norris}, {Schnitzeler}, {Subrahmanyan}, {Urquhart}, {Thompson}, \& {Wark}}]{2011MNRAS.416..832W}
{Wilson}, W.~E., {Ferris}, R.~H., {Axtens}, P., {et~al.} 2011, \mnras, 416, 832, \dodoi{10.1111/j.1365-2966.2011.19054.x}

\bibitem[{{Wright} {et~al.}(2010){Wright}, {Eisenhardt}, {Mainzer}, {Ressler}, {Cutri}, {Jarrett}, {Kirkpatrick}, {Padgett}, {McMillan}, {Skrutskie}, {Stanford}, {Cohen}, {Walker}, {Mather}, {Leisawitz}, {Gautier}, {McLean}, {Benford}, {Lonsdale}, {Blain}, {Mendez}, {Irace}, {Duval}, {Liu}, {Royer}, {Heinrichsen}, {Howard}, {Shannon}, {Kendall}, {Walsh}, {Larsen}, {Cardon}, {Schick}, {Schwalm}, {Abid}, {Fabinsky}, {Naes}, \& {Tsai}}]{2010AJ....140.1868W}
{Wright}, E.~L., {Eisenhardt}, P. R.~M., {Mainzer}, A.~K., {et~al.} 2010, \aj, 140, 1868, \dodoi{10.1088/0004-6256/140/6/1868}

\bibitem[{{Xie} {et~al.}(2022){Xie}, {Di Marco}, {La Monaca}, {Liu}, {Muleri}, {Bucciantini}, {Romani}, {Costa}, {Rankin}, {Soffitta}, {Bachetti}, {Di Lalla}, {Fabiani}, {Ferrazzoli}, {Gunji}, {Latronico}, {Negro}, {Omodei}, {Pilia}, {Trois}, {Watanabe}, {Agudo}, {Antonelli}, {Baldini}, {Baumgartner}, {Bellazzini}, {Bianchi}, {Bongiorno}, {Bonino}, {Brez}, {Capitanio}, {Castellano}, {Cavazzuti}, {Ciprini}, {De Rosa}, {Del Monte}, {Di Gesu}, {Donnarumma}, {Doroshenko}, {Dov{\v{c}}iak}, {Ehlert}, {Enoto}, {Evangelista}, {Garcia}, {Hayashida}, {Heyl}, {Iwakiri}, {Jorstad}, {Karas}, {Kitaguchi}, {Kolodziejczak}, {Krawczynski}, {Liodakis}, {Maldera}, {Manfreda}, {Marin}, {Marinucci}, {Marscher}, {Marshall}, {Massaro}, {Matt}, {Mitsuishi}, {Mizuno}, {Ng}, {O'Dell}, {Oppedisano}, {Papitto}, {Pavlov}, {Peirson}, {Perri}, {Pesce-Rollins}, {Petrucci}, {Possenti}, {Poutanen}, {Puccetti}, {Ramsey}, {Ratheesh}, {Sgr{\'o}}, {Slane}, {Spandre}, {Tamagawa}, {Tavecchio}, {Taverna}, {Tawara}, {Tennant}, {Thomas}, {Tombesi},
  {Tsygankov}, {Turolla}, {Vink}, {Weisskopf}, {Wu}, \& {Zane}}]{2022Natur.612..658X}
{Xie}, F., {Di Marco}, A., {La Monaca}, F., {et~al.} 2022, \nat, 612, 658, \dodoi{10.1038/s41586-022-05476-5}

\end{thebibliography}
\bibliographystyle{aasjournal}
\end{document}